\documentclass{cernrep}

\usepackage{graphicx}
\usepackage{epsfig}
\usepackage{subfigure,rotating,rotate}

\def\ga{\mathrel{\rlap{\raise.6ex\hbox{$>$}}{\lower.6ex\hbox{$\sim$}}}}
\def\la{\mathrel{\rlap{\raise.6ex\hbox{$<$}}{\lower.6ex\hbox{$\sim$}}}}

\newcommand{\micron}{\ensuremath{\mathrm{\;\mu m}}\xspace}
\newcommand{\cm}{\ensuremath{\mathrm{\; cm}}\xspace}
\newcommand{\mm}{\ensuremath{\mathrm{\; mm}}\xspace}
\newcommand{\mus}{\ensuremath{\mathrm{\;\mu s}}\xspace}
\newcommand{\keV}{\ensuremath{\mathrm{\; keV}}\xspace}
\newcommand{\MeV}{\ensuremath{\mathrm{\; MeV}}\xspace}
\newcommand{\GeV}{\ensuremath{\mathrm{\; GeV}}\xspace}
\newcommand{\TeV}{\ensuremath{\mathrm{\; TeV}}\xspace}

\newcommand{\Bz}{\ensuremath{B^{0}}\xspace}

\newcommand{\Hz}{\ensuremath{H^{0}}\xspace}

\newcommand{\rpv}{\ensuremath{\rlap{\kern.2em/}R}\xspace}

\input{babarsym.tex}

\renewcommand{\Re}{{\rm Re}}
\renewcommand{\Im}{{\rm Im}}

\begin{document}
\begin{flushright}
CERN-PH-TH/2007-162\\
SLAC-PUB-12817\\
KEK-TH-1173\\
LAL 07-175\\[2cm]
\end{flushright}

{\huge\bf On the Physics Case of a Super Flavour Factory}

\vspace{1cm}

{\large\bf T.Browder,$^1$ M.Ciuchini,$^2$ T.Gershon,$^3$ M.Hazumi,$^{4,7}$ T.Hurth,$^{5, 6}$ Y.Okada,$^{4,7}$ A.Stocchi$^{8,9}$}  

{\large\it $^1$ University of Hawaii at Manoa, Honolulu, Hawaii 96822, USA}

{\large\it $^2$ INFN Sezione di Roma Tre and Dip. di Fisica, Univ. of Roma Tre, I-00146 Rome, Italy}

{\large\it $^3$ University of Warwick, Department of Physics, Coventry CV4 7AL, UK} 

{\large\it $^4$  High Energy Accelerator Research Organization (KEK), Tsukuba, Japan} 

{\large\it $^5$ CERN, Department of Physics, Theory Unit, CH-1211 Geneva 23, Switzerland} 

{\large\it $^6$ SLAC, Stanford University, Stanford, CA 94309, USA}

{\large\it $^7$ Graduate University for Advanced Studies (Sokendai), Tsukuba, Japan}  

{\large\it $^8$ Laboratoire de l'Accelerateur Lineaire IN2P3-CNRS, France}

{\large\it $^9$  University  de Paris-Sud, BP34, F-91898 Orsay cedex, France}

\vspace{1cm} 

{\large\bf Abstract}\\ 
We summarize the physics case of a high-luminosity $e^+e^-$ flavour 
factory  collecting an integrated luminosity of $50-75 \ {\rm ab}^{-1}$.  
Many New Physics sensitive measurements involving $B$ and $D$ mesons 
and $\tau$ leptons, unique to a Super Flavour Factory, 
can be performed with excellent sensitivity to new particles with 
masses up to  $\sim 100$ (or even $\sim 1000$) TeV.
Flavour- and $\CP$-violating couplings of new particles 
that may be discovered at the LHC
can be measured in most scenarios, 
even in unfavourable cases assuming minimal flavour violation.  
Together with the LHC, a Super Flavour Factory, 
following either the SuperKEKB or the Super$B$ proposal, 
could be soon starting the 
project of reconstructing  the New Physics Lagrangian.

\newpage 

\section{Introduction}

Many open fundamental questions of particle physics are related to flavour: How
many families are there? What is their origin? How are neutrino and quark masses
and mixing angles generated? Do there exist new sources of flavour and $\CP$
violation beyond those we already know? What is the relation between the flavour
structure in the lepton and quark sectors? Future flavour experiments will
attempt to address these questions providing the exciting possibility to learn
something about physics at energy scales much higher than those reachable by
current experiments.

The Standard Model (SM) of elementary particles has been very successful in
explaining a wide variety of existing experimental data. It accounts for a range
of phenomena from low-energy physics (less than a GeV), such as kaon decays, to
high-energy (a few hundred GeV) processes involving real weak gauge bosons ($W$
and $Z$) and top quarks. There is, therefore, little doubt that the SM is the
theory to describe physics below the energy scale of several hundred GeV, namely
all that has been explored so far.

In spite of the tremendous success of the SM, it is fair to say that the flavour
sector of the SM is much less understood than its gauge sector, reflecting our
lack of answers to the questions mentioned above. Masses and mixing of the
quarks and leptons, which have a significant but unexplained hierarchy pattern,
enter as free parameters to be determined experimentally. In fact, while
symmetries shape the gauge sector, no principle governs the flavour structure of
the SM Lagrangian. Yukawa interactions provide a phenomenological description of
the flavour processes which, while successful so far, leaves most fundamental
questions unanswered. Hence the need to go beyond the SM.

Indeed the search for evidence of physics beyond the SM is the main goal of
particle physics in the next decades. The LHC at CERN will start soon looking
for the Higgs boson, the last missing building block of the SM. At the same time
it will intensively search for New Physics (NP), for which there are solid
theoretical motivations related to the quantum stabilization of the Fermi scale
to expect an appearance at energies around $1$ TeV.

However, pushing the high-energy frontier, {i.e.} increasing the available
centre-of-mass energy in order to produce and observe new particles, is not the
only way to look for NP. New particles could reveal themselves through their
virtual effects in processes involving only  standard particles 
as has been the case several  times in the history of particle physics. 
For these kind of searches
the production thresholds are not an issue.
Since quantum effects become typically smaller as the mass of the virtual
particles increases, the name of the game is rather high precision. 
As a matter of fact, high-precision measurements 
probe NP energy scales
inaccessible at present and next-generation colliders at the energy frontier.

Flavour physics is the best candidate as a tool for NP searches 
through quantum effects for several reasons. 
Flavour Changing Neutral Currents (FCNC), neutral
meson-antimeson mixing and $\CP$ violation occur at the loop level in the SM and
therefore are potentially subject to ${\cal O}(1)$ NP virtual corrections. In
addition, quark flavour violation in the SM is governed by the weak interaction
and suppressed by the small quark mixing angles. Both these features are not
necessarily shared by NP which, in such cases, could produce very large effects.
Indeed, the inclusion in the SM of generic NP flavour-violating terms with
natural ${\cal O}(1)$ couplings is known to violate present experimental
constraints unless the NP scale is pushed up to $10$--$100$ TeV depending on the
flavour sector. This difference between the NP scale emerging from flavour
physics and the one suggested by Higgs physics could be a problem for model
builders (the so-called flavour problem), but it clearly indicates that flavour
physics has the potential to push the explored NP scale in the $100$ TeV region.
On the other hand, if the NP scale is indeed close to $1$ TeV, the flavour
structure of NP must  be highly non-trivial and the experimental determination
of the flavour-violating couplings is particularly interesting.

Let us elaborate on this latter option. Any new-physics model, established at
the TeV scale to solve the gauge hierarchy problem, includes new flavoured
particles and new flavour- and $\CP$-violating parameters. Therefore, such a
model must provide a solution also to the flavour and $\CP$ problems, namely how
new flavour changing neutral currents and $\CP$-violating phenomena are
suppressed. This may be related to other interesting questions. For instance, in
supersymmetry the flavour problem is directly linked to the crucial issue of
supersymmetry breaking. Similar problems also occur in models of
extra-dimensions (flavour properties of Kaluza-Klein states), Technicolour models
(flavour couplings of Techni-fermions), little-Higgs models (flavour couplings
of new gauge bosons and fermions) and multi-Higgs models ($\CP$-violating Higgs
couplings). Once NP is found at the TeV scale, precision measurements of
flavour- and $\CP$-violating observables would 
shed light on the detailed structure of the underlying model.

On quite general grounds, 
quantum effects in flavour processes explore a parameter space including the
NP scale and the NP flavour- and $\CP$-violating couplings. In specific models
these are related to fundamental parameters such as masses and couplings of new
particles. In particular, NP effects tend to disappear at large NP scales as 
well as for small couplings. Therefore a crucial question is: could NP be
flavour-blind, thus making searches for it  with flavour physics unfeasible?
Fortunately, the concept of Minimal Flavour Violation (MFV) provides a negative
answer: even if NP does  not contain new sources of flavour and $\CP$ violation,
the flavour-violating couplings present in the SM are enough to produce a new
phenomenology that  makes flavour processes sensitive to the presence of new
particles. In other words, MFV puts a lower bound on the flavour effects
generated by NP appearing at a given mass scale, a sort of ``worst case''
scenario for the flavour-violating couplings extremely useful to exclude NP
flavour-blindness and assess the ``minimum'' performance of flavour physics in
searching for NP, always keeping in mind that larger effects are quite possible
and easily produced in many scenarios beyond MFV.

In the light of the above considerations, a Super Flavour Factory (SFF), 
following the recent SuperKEKB~\cite{Hashimoto:2004sm} and 
Super$B$~\cite{Bona:2007qt} proposals, has one mission: 
to search for new physics in the flavour sector 
exploiting a huge leap in integrated luminosity
and the wide range of observables that it can measure.
However this goal can be pursued in different ways depending on whether
evidence of NP has been found at the time a SFF starts taking data.

In either scenario, a SFF can search for evidence of NP irrespective of the
values of the new particle masses and of the unknown flavour-violating
couplings. A large number of measurements could provide evidence for NP 
at a SFF. 
A first set is given by measurements of observables which are predicted 
by the SM with small uncertainty, including those which are 
vanishingly small (the so-called null tests). 
Among them are the flavour-violating $\tau$ decays,
direct $\CP$ asymmetries in $B\to X_{s+d}\gamma$, 
in $\tau$ decays and in some non-leptonic $D$ decays, 
$\CP$ violation in neutral charm meson mixing,
the dilepton invariant mass at which the
forward-backward asymmetry of $B\to X_s\ell^+\ell^-$ vanishes, 
and lepton universality violating $B$ and $\tau$ decays. 
Any deviation, as small as a SFF could measure, 
from its SM value of any observable in this
set could be ascribed to NP with 
essentially no uncertainty. 
A second set of NP-sensitive observables, 
including very interesting decays such as 
$b\to s$ penguin-dominated non-leptonic $B$ decays, $B\to \tau\nu$, 
$B \to D^{(*)}\tau\nu$, $B\to K^*\gamma$, $B\to \rho\gamma$, and many others, 
require more accurate determinations of SM contributions 
and improved control of the hadronic uncertainties with respect 
to what we can do today in order to match the experimental precision 
achievable at a SFF and to allow for an unambiguous
identification of a NP signal. 
The error on the SM can be reduced using the improved determination of 
the Cabibbo-Kobayashi-Maskawa (CKM) matrix provided by a SFF itself.
This can be achieved using generalized CKM fits which allow for
a $1\%$ determination of the CKM parameters using tree-level 
and $\Delta F=2$ processes even in the presence of generic NP contributions. 
As far as hadronic uncertainties are concerned, 
the extrapolation of our present knowledge
and techniques shows that it is possible to reach the required accuracy by the
time a SFF will be running using improved lattice QCD results obtained with
next-generation computers~\cite{Bona:2007qt} 
and/or bounding the theoretical uncertainties with
data-driven methods exploiting the huge SFF data sample.

As we already noted, the NP search at a  SFF could reveal the virtual effect of
particles with masses of hundreds of TeV and in some cases, notably $\Delta F=2$
processes, even thousands of TeV depending on the values of the
flavour-violating couplings. 
Therefore this search is worth doing 
irrespective of whether NP has already been found or not. 
If new particles are discovered at the energy frontier, 
a SFF could enlarge the spectrum 
providing evidence of heavier states not accessible otherwise; 
if not, quantum effects measurable at a SFF
could be the only option to look for NP for a long time.

If the LHC finds NP at the TeV scale --
in particular if the findings include one (or more) new flavoured particle(s) --
then a SFF could measure its flavour- and $\CP$-violating couplings. 
Indeed all terms of the NP Lagrangian non-diagonal in
the flavour space are barely  accessible at the LHC. A SFF would be needed to
accomplish the task of reconstructing them. It seems able to do that even in
the unfavourable cases provided by most MFV models. Indeed, for the purpose of
inferring the NP Lagrangian from experiments, the LHC and SFF physics programmes are
complementary. 

Finally, 
it must be emphasised that while a Super {\it Flavour} Factory 
will perform detailed studies of beauty, charm and tau lepton physics,
the results will be highly complementary to those on several 
important observables related to $B_s$ meson oscillations, 
kaon and muon decays that will be measured elsewhere.
Most benchmark charm measurements, 
in particular interesting NP-related measurements such as
$\CP$ violation in charm mixing,
  will still  be statistics-limited after the 
  CLEO$c$, BESIII and $B$ factory  projects are completed, 
  and can only be pursued to their ultimate precision at a SFF.
  Operation at the $\Upsilon(5{\rm S})$ resonance
  provides the possibility of exploiting the clean $e^+e^-$ environment
  to measure $B^0_s$ decays with neutral particles in the final state,
  which will  complement the channels that can be measured at LHCb.
  A SFF has sensitivity for $\tau$ physics that is far superior 
  to any other existing or proposed experiment,
  and the physics reach can be extended even further by
  the possibility to operate with polarized beams.
  It is particularly noteworthy that 
  the combined information on $\mu$ and $\tau$ flavour violating decays 
  that will be provided by MEG~\cite{Ritt:2006cg} together with a SFF
  can shed light on the mechanism responsible for lepton flavour violation.

\section{Experimental Sensitivities}

A Super Flavour Factory (SFF) with integrated luminosity 
of $50$--$75 \ {\rm ab}^{-1}$
can perform a wide range of important measurements and dramatically improve 
upon the results from the current generation of $B$ Factories.
Many of these measurements cannot be made in a hadronic environment,
and are unique to a SFF. 
The experimental sensitivities of a SFF  can be 
schematically classified in two categories:

\begin{itemize}
\item {\it Searching for New Physics:} \\
  Many of the measurements that can be made at a SFF 
  are highly sensitive to NP  effects,
  and those with precise SM  predictions are potential 
  discovery channels.
  As an example: the mixing-induced $\CP$ asymmetry parameter
  for $B^0 \to \phi K^0$ decays 
  can be measured to a precision of $0.02$,
  as can equivalent parameters for numerous hadronic decay channels
  dominated by the $b \to s$ penguin transition.
  These constitute very stringent tests of any NP scenario
  which introduces new $\CP$ violation sources, beyond the Standard Model.
     The presence of new sources of $\CP$ violation 
    in $D^0$--$\bar{D}^0$ mixing, where the SM background is negligible,
    can be tested to similar precision.
    New physics that appears in the $D^0$ sector (involving up-type quarks) 
    may be different or complementary to that in the $B^0_d$ or $B^0_s$ 
sectors.
  Direct $\CP$ asymmetries can be measured to the 
  fraction of a percent level in $b \to s \gamma$ decays,
  using both inclusive and exclusive channels,
  and $b \to s \ell^+\ell^-$ can be equally thoroughly explored.
  Equally precise searches for direct $\CP$ violation in 
    charm or $\tau$ decays provide additional NP sensitivity,
    since the SM background is largely absent.
  At the same time, a SFF
  can access channels that are sensitive to NP
  even when there are no new sources of $\CP$ violation,
  such as the photon polarization in $b \to s\gamma$,
  and the branching fractions of $B^+ \to \ell^+ \nu_\ell$,
  the latter being sensitive probes of NP in MFV scenarios
  with large $\tan \beta$. 
  Furthermore, rare FCNC decays of the $\tau$ lepton are particularly
  interesting since lepton flavour violation sources involving the third 
  generation are naturally the largest.  
  Any of these measurements constitutes clear motivation for a SFF.
\item {\it Future metrology of the CKM matrix:} \\
  There are several measurements that are unaffected by NP
  in many likely scenarios, and which allow the extraction
  of the CKM parameters even in the presence of such NP effects.
  Among these, the angle $\gamma$ can be measured with a precision
  of $1$--$2{^\circ}$,
  where the precision is limited only by statistics,
  not by systematics or by theoretical errors.
  By contrast,
  the determination of the elements $|V_{ub}|$ and $|V_{cb}|$
  will be limited by theory,
  but the large data sample of a SFF  will allow
  many of the theoretical errors to be much improved.
  With anticipated improvements in lattice QCD calculations,
  the precision on $|V_{ub}|$ and $|V_{cb}|$ can be driven
  down to the percent level.
  These measurements could allow
  tests of the consistency of the Standard Model at a few per mille level and provide the
  NP phenomenological analyses with a determination of the CKM
  matrix at the percent level.
\end{itemize}

In Table~\ref{tab:superb} we give indicative estimates of the precision 
on some of the most important observables that can be achieved by a 
SFF with integrated luminosity of $50$--$75 \ {\rm ab}^{-1}$.
Here we have not attempted to comment on  the whole range of measurements 
that can be performed by such a machine,
but instead focus on channels with the greatest phenomenological impact.
For more details, including a wide range of additional measurements, 
we guide the reader to the reports~\cite{Hashimoto:2004sm,Bona:2007qt,Akeroyd:2004mj,Hewett:2004tv,superKEKB}, where also all original references are 
given.

\begin{table}[!ht]
  \begin{center}
    \caption{Expected sensitivity that can be achieved 
      on some of the most important observables,
      by a SFF with integrated luminosity of 
      $50$--$75 \ {\rm ab}^{-1}$. The range of values given
allow for possible variation in the total integrated luminosity, in the
accelerator and detector design, and in limiting systematic effects.
      For further details, refer to~\cite{Bona:2007qt,superKEKB}.
    }
    \label{tab:superb}    
    \begin{tabular}{l@{\hspace{15mm}}c}
      \hline \hline
      Observable                      & Super Flavour Factory  sensitivity \\
      \hline
      $\sin(2\beta)$ ($J/\psi\,K^0$)        &  0.005--0.012      \\
      $\gamma$ ($B \to D^{(*)}K^{(*)}$)       &  $1$--$2^\circ$  \\
      $\alpha$ ($B \to \pi\pi, \rho\rho, \rho\pi$)  &  $1$--$2^\circ$    \\
      $\left| V_{ub} \right|$ (exclusive)   &  $3$--$5\%$   \\
      $\left| V_{ub} \right|$ (inclusive)   &  $2$--$6\%$   \\
      \hline
      $\bar{\rho}$                          & $1.7$--$3.4\%$  \\
      $\bar{\eta}$                          & $0.7$--$1.7\%$  \\
      \hline
      $S(\phi K^0)$                         &  0.02--0.03        \\
      $S(\eta^\prime K^0)$                  &  0.01--0.02        \\
      $S(\KS\KS\KS)$                        &  0.02--0.04        \\
      \hline
      $\phi_D$                              & $1$--$3^\circ$      \\
      \hline
      $\BR(B \to \tau \nu)$                 &  $3$--$4\%$         \\
      $\BR(B \to \mu \nu)$                  &  $5$--$6\%$         \\
      $\BR(B \to D \tau \nu)$               &  $2$--$2.5\%$       \\
      \hline
      $\BR(B \to \rho \gamma)/\BR(B \to K^* \gamma)$ &  $3$--$4\%$ \\
      $A_{CP}(b \to s \gamma)$              &  $0.004$--$0.005$   \\
       $A_{CP}(b \to (s+d) \gamma)$          &  $0.01$             \\
      $S(\KS\pi^0\gamma)$                   &  $0.02$--$0.03$     \\
      $S(\rho^0\gamma)$                     &  $0.08$--$0.12$        \\
      $A^{\rm FB}(B \to X_s \ell^+ \ell^-) \ s_0$ &  $4$--$6\%$              \\
      $\BR(B \to K \nu \bar{\nu})$     &  $16$--$20\%$             \\
      \hline
      $\BR(\tau \to \mu\gamma)$         & $2$--$8 \times 10^{-9}$  \\
      $\BR(\tau \to \mu\mu\mu)$     & $0.2$--$1 \times 10^{-9}$  \\
      $\BR(\tau \to \mu\eta)$           & $0.4$--$4 \times 10^{-9}$  \\
      \hline \hline
    \end{tabular}
  \end{center}
\end{table}

{\it The most important measurements within the CKM metrology are} 
the angles of the Unitarity Triangle, 
the angle $\beta$ (also known as $\phi_1$), 
measured using mixing-induced $\CP$ violation in $B^0 \to J/\psi\,K^0$, 
the angle $\alpha$ ($\phi_2$),
measured using rates and asymmetries in 
$B \to \pi\pi$~\footnote{Notice that this method for extracting alpha is insensitive to NP in QCD penguins. However it could be affected by isospin-breaking NP contributions.}, $\rho\pi$ and $\rho\rho$, 
and the angle $\gamma$ ($\phi_3$),
measured using rates and asymmetries in $B \to D^{(*)}K^{(*)}$ decays,
using final states accessible to both $D^0$ and $\bar{D}^0$. 
Moreover, a SFF will improve our knowledge of the lengths  
of the sides of the Unitarity Triangle.
In particular, the CKM matrix element $\left| V_{ub} \right|$
will be precisely measured through both inclusive and exclusive
semileptonic $b \to u$ decays.

{\it Among the measurements sensitive for New Physics,} 
there are the mixing-induced $\CP$ violation parameters 
in charmless hadronic $B$ decays dominated by the $b \to s$ penguin transition,
$S(\phi K^0)$, $S(\eta^\prime K^0)$ and $S(K^0_S K^0_S K^0_S)$.
Within the Standard Model these give the same value of $\sin(2\beta)$
that is determined in $B^0 \to J/\psi\,K^0$ decays,
up to a level of theoretical uncertainty that is 
estimated to be $\sim 2$--$5\%$ within factorization.
(The theoretical error in these and other modes, such as $B\to K_S \pi^0$, 
can be also bounded with data-driven methods~\cite{Silvestrini:2007yf}. 
Presently these give larger uncertainties but will become more precise
as more data is available.)
Many extensions of the Standard Model result in deviations 
from this prediction. 
Another distinctive probe of new sources of $\CP$ violation is $\phi_D$, 
the $\CP$ violating phase in neutral $D$ meson mixing,
which is negligible in the SM and can be precisely measured using,
for example, $D \to \KS\pi^+\pi^-$ decays.
Furthermore, branching fractions for leptonic and semileptonic $B$ decays 
are sensitive to charged Higgs exchange.
In particular these modes are sensitive to new physics,
even in the unfavourable minimal flavour violation scenario,
with a large ratio of the Higgs vacuum expectation values, $\tan \beta$. 
Measurements of rare radiative and electroweak penguin processes
are well-known to be particularly  sensitive to new physics:
The ratio of branching fractions
$\BR(B \to \rho \gamma)/\BR(B \to K^* \gamma)$
depends on the ratio of CKM matrix parameters
$\left| V_{td} / V_{ts} \right|$,
with additional input from lattice QCD.
Within the Standard Model this result must be consistent with 
constraints from the Unitarity Triangle fits.
The inclusive $\CP$ asymmetries $A_{CP}(b \to s \gamma)$
or $A_{CP}(b \to (s+d)  \gamma)$
are predicted in the Standard Model to be small or exactly zero respectively
with well understood theoretical uncertainties.
The mixing-induced $\CP$ asymmetry in radiative $b \to s$ transitions,
measured for example through $S(K^0_S \pi^0\gamma)$,
is sensitive to the emitted photon polarization.
Within the SM the photon is strongly polarized,
and the mixing-induced asymmetry small,
but new right-handed currents can break this prediction
even without the introduction of any new $\CP$ violating phase.
Similarly, $S(\rho^0\gamma)$ probes radiative $b \to d$ transitions.
The dilepton invariant mass squared $s$ 
at which the forward-backward asymmetry 
in the distribution of $B \to X_s \ell^+ \ell^-$ decays is zero
(denoted $A^{\rm FB}(B \to X_s \ell^+ \ell^-) \ s_0$),
for which the theoretical uncertainty 
of the Standard Model prediction is small,
is sensitive to NP in electroweak penguin operators; 
finally, the branching fraction for the rare electroweak penguin decay 
$B \to K \nu \bar{\nu}$ is an important probe for NP  
even if this appears only well above the electroweak scale. 
A SFF also allows for the 
measurement of branching ratios of lepton flavour violating $\tau$ decays, 
such as  $\tau \to \mu\gamma$, $\tau \to \mu\mu\mu$ and $\tau \to \mu\eta$.
Within the Standard Model, these are negligibly small,
but many models of new physics create observable lepton flavour
violation signatures.

For some of the entries of Table~\ref{tab:superb}  
some additional comments are in order: 
\begin{itemize}
\item  
  With such large data samples as will be accumulated by a SFF,
  the uncertainty on several measurements 
  will be dominated by systematic errors.
  Estimating the ultimate precision therefore requires some knowledge
  of how these systematic uncertainties can be improved.
  One such important channel is the mixing-induced $\CP$ asymmetry
  in $B^0 \to J/\psi\,K^0$, which measures $\sin(2\beta)$ in the SM.
  The systematic uncertainties in the current $B$ factory analyses
  are around $1$--$2\%$, 
  coming mainly from uncertainties in the vertex detector alignment 
  and beam spot position.
  Another example is direct $\CP$ asymmetry,
  both in exclusive and inclusive modes.
  Measurements with precision better than $1\%$ require knowledge
  of detector asymmetries at the same level.
  Reduction of these errors will be highly challenging,
  but there is some hope that improvement 
  by a factor of about two may be possible. 
\item 
  The precision that can be achieved on $\left| V_{ub} \right|$
  depends on improvements in the theoretical treatment.
  The most notable effect is for the exclusive channels,
  where reduction of the error on form factors calculated in 
  lattice QCD  is extremely important.
\item
    The sensitivities for some measurements depend on hadronic parameters
    that are not yet well known.
    For example, for $\phi_D$ to be measured at least one of the 
    $D^0$--$\bar{D}^0$ mixing parameters $x_D$ and $y_D$ must be nonzero.
    The first evidence for charm mixing has recently been reported~\cite{Staric:2007dt,Aubert:2007wf},
    but large ranges for the obtained parameters are still allowed.
    Our estimate of the sensitivity is obtained by extrapolating 
    results from the $D\to K_S \pi^+\pi^-$ time-dependent analysis~\cite{Abe:2007rd},
    which currently appears to be the single most sensitive channel,
    although better constraints can certainly be obtained by combining
    information from multiple decays modes.
\item 
  The specific details of the accelerator and detector configuration
  are important considerations for some measurements.
  For studies of mixing-induced $\CP$ asymmetry
  that obtain the $B$ decay vertex position from a reconstructed $\KS$ meson
  (such as $B^0\to\KS\KS\KS$ and $B^0\to\KS\pi^0\gamma$)
  the geometry of the vertex detector plays an important role --
  better precision is achieved for a larger vertex detector.
  Similarly, several channels with missing energy
  (such as $B\to\tau\nu_\tau$, $B \to D\tau\nu_\tau$ and $B\to K\nu\bar{\nu}$)
  make full use of the constraints available in 
  $\Upsilon(4{\rm S}) \to B\bar{B}$ decays
  by fully reconstructing one $B$ meson to know the kinematics of the other.
  Such measurements are dependent on the background condition 
  and the hermeticity of the detector.  
  Indeed, it is obvious that the sensitivity for all measurements 
  depends strongly on the detector performance,
  and improvements in, {\it e.g.}, vertexing and particle identification
  capability will be of great benefit to separate signal from background.
\item 
  The sensitivity to very rare processes,
  such as the lepton flavour violating decay $\tau \to \mu\gamma$
  depends strongly on how effectively the background may be reduced
  and on other possible improvements to the analysis techniques used.
\end{itemize}

\begin{figure}[!ht]
  \begin{center}
    \includegraphics[width=9cm]{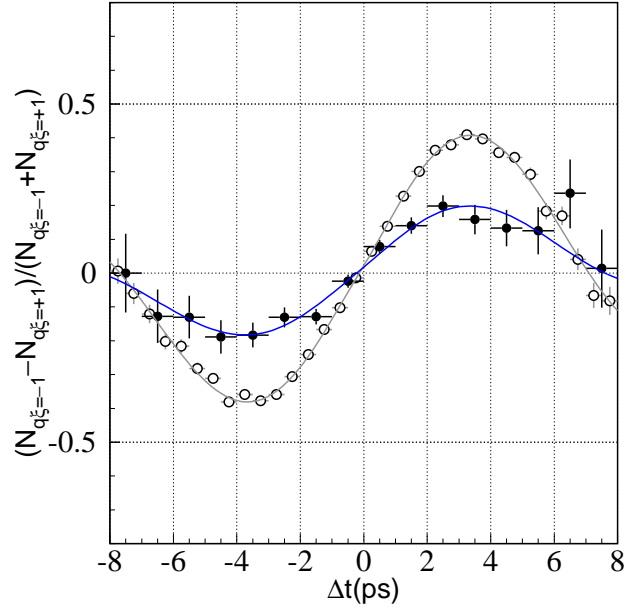}
    \caption{
      Simulation of new physics effects in $B^0 \to \phi K^0$,
      as could be observed by a SFF.  
      The open circles show simulated $B^0 \to J/\psi\,K^0$ events,
      the filled circles show simulated $B^0 \to \phi K^0$ events.
      Both have curves showing fit results superimposed.
      From~\cite{superKEKB}.
    }
    \label{fig:phiKs_data}
  \end{center}
\end{figure}

{\it The sensitivities  of these measurements to New Physics effects} 
may be shown  by a few examples:
In Figure~\ref{fig:phiKs_data} we show a simulation of the 
time-dependent asymmetry in $B^0 \to \phi K^0$,
compared to that for $B^0 \to J/\psi\,K^0$.
The events are generated using the current central values of the measurements.
With the precision of a SFF and the present central values, the difference between the two data sets is larger than the theoretical expectation, showing evidence of NP contributions.

\begin{figure}[!ht]
  \begin{center}
    \includegraphics[width=9cm]{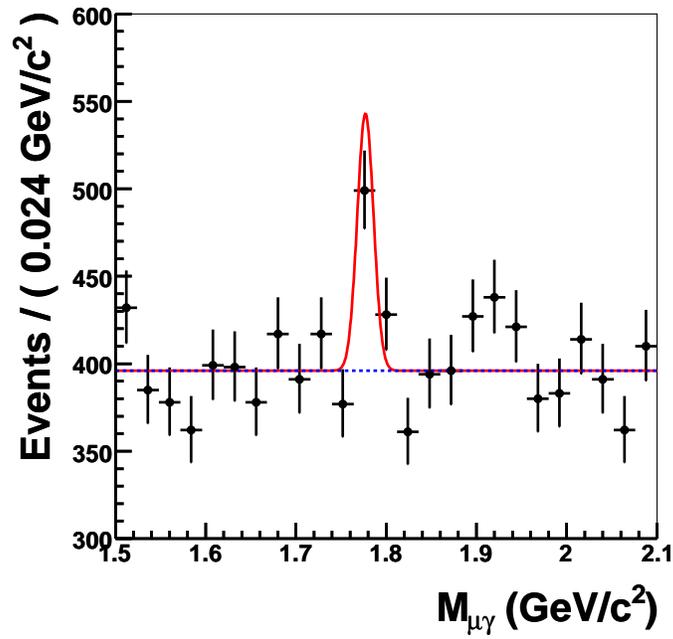}
    \caption{
      Monte Carlo simulation of the appearance of $\tau \to \mu \gamma$
      at a SFF.
      A clear peak in the $\mu\gamma$ invariant mass distribution 
      is visible above the background.
      The branching fraction used in the simulation is
      $\BR(\tau \to \mu \gamma) = 10^{-8}$,
      an order of magnitude below the current upper limit.
      With $75 \ {\rm ab}^{-1}$ of data the significance
      of such a decay is expected to exceed $5\sigma$.
    }
    \label{fig:tauLFV_data}
  \end{center}
\end{figure}

In Figure~\ref{fig:tauLFV_data} we show how lepton flavour violation
in the decay $\tau \to \mu \gamma$ may be discovered at a SFF.
The simulation corresponds to a branching fraction of 
$\BR(\tau \to \mu \gamma) = 10^{-8}$,
which is within the range predicted by many new physics models.
The signal is clearly observable, 
and well within the reach of a SFF.
The simulation includes the effects of irreducible background 
from initial state radiation photons,
though improvements in the detector and in the analysis 
may lead to better control of this limitation.
Other lepton flavour violating decay modes, 
such as $\tau \to \mu\mu\mu$ do not suffer from this background,
and have correspondingly cleaner experimental signatures.

The differences between the SFF physics programme and those of the 
current $B$ factories are striking.
At a SFF measurements of known rare processes such as 
$b\to s \gamma$ or $\CP$ violation in hadronic $b\to s$ penguin transitions
such as $B^0 \to \phi \KS$ will be advanced to unprecedented precision.
Channels which are just being observed in the existing data,
such as $B^0 \to \rho^0\gamma$, $B^+ \to \tau^+\nu_\tau$ 
and $B \to D^{(*)}\tau\nu$ will become precision measurements at a SFF.
Furthermore, detailed studies of decay distributions and asymmetries
that cannot be performed with the present statistics,
will enable the sensitivity to NP to be significantly improved.
Another salient example lies in $D^0$--$\bar{D}^0$ oscillations:
the current evidence for charm mixing, which cannot be interpreted
in terms of New Physics, opens the door for precise measurements of 
the $\CP$ violating phase in charm mixing, which is known to 
be zero in the Standard Model with negligible uncertainty.

In addition, these measurements will be accompanied by 
dramatic discoveries of new modes and processes. 
These will include decays such as $B \to K \nu \bar{\nu}$, 
which is the signature of the theoretically clean quark level process 
$b \to s \nu \bar{\nu}$. 
The high statistics and 
clean environment of a SFF allow for the accompanying $B$ meson
to be fully reconstructed in a hadronic decay mode, which then
in turn allows a one-charged prong rare decay to be isolated. 
Another example is $B^+ \to \pi^+ \ell^+ \ell^-$, 
the most accessible $b\to d \ell^+\ell^-$ process. 
These decays are the next level beyond $b\to s \ell^+\ell^-$ decays, 
which were first observed in the $B$ factory era. 
Such significant advances will result in a strong phenomenological impact
of the Super Flavour Factory physics programme.

{\it Comparison with LHCb:} 
Since a SFF will take data in the LHC era,
it is reasonable to ask how the physics reach
compares with the $B$ physics potential of the LHC experiments,
most notably LHCb.
By 2014, the LHCb experiment is expected to have accumulated
$10 \ {\rm fb}^{-1}$ of data
from $pp$ collisions at a luminosity of
$\sim 2 \times 10^{32} \ {\rm cm}^{-2} {\rm s}^{-1}$.
In the following we assume the most recent estimates of LHCb 
sensitivity with that data set~\cite{schneider}.
Note that LHCb is planning an upgrade where they would run
at 10 times the initial design luminosity and record a data sample of 
about $100 \ {\rm fb}^{-1}$~\cite{wilkinson}.

The most striking outcome of any comparison between SFF and LHCb
is that the strengths of the two experiments are largely complementary.
For example, the large boost of the $B$ hadrons produced at LHCb 
allows studies of the oscillations of $B_s$ mesons
while many of the measurements that constitute the primary
physics motivation for a SFF cannot be performed
in the hadronic environment, including rare decay modes with missing energy
such as $B^+ \to \ell^+\nu_\ell$ and $B^+ \to K^+\nu\bar{\nu}$. 
Measurements of the CKM matrix elements $|V_{ub}|$ and $|V_{cb}|$
and inclusive analyses of processes such as $b \to s\gamma$
also benefit greatly from the SFF environment.
At LHCb the reconstruction efficiencies are reduced 
for channels containing several neutral particles and 
for studies where 
the $B$ decay vertex must be determined from a $K^0_S$ meson. 
Consequently, a SFF has unique potential to measure the photon polarization
via mixing-induced $\CP$ violation in $B^0 \to K^0_S \pi^0 \gamma$.
Similarly, a SFF is well placed to study possible NP effects in
hadronic $b \to s$ penguin decays as it 
can measure precisely the $\CP$ asymmetries in many $B^0_d$ decay modes 
including $\phi K^0$, $\eta^\prime K^0$, $K^0_S K^0_S K^0_S$ or 
$K^0_S \pi^0$. While LHCb will have limited capability for these channels,
it can achieve complementary measurements
using decay modes such as $B^0_s \to \phi \gamma$ and $B^0_s \to \phi\phi$
for radiative and hadronic $b \to s$ transitions respectively.

Where there is overlap,
the strength of the SFF programme in its ability to use multiple 
approaches to reach the objective becomes apparent.
For example, LHCb will be able to measure
$\alpha$ to about $5^\circ$ precision using $B \to \rho\pi$,
but would not be able to access the full information in the
$\pi\pi$ and $\rho\rho$ channels, which is necessary to drive
the uncertainty down to the $1$--$2^\circ$ level of a SFF.
Similarly, LHCb can certainly measure $\sin(2\beta)$
through mixing-induced $\CP$ violation in $B^0 \to J/\psi K^0_S$ decay
to high accuracy (about 0.01),
but will have less sensitivity to make the complementary measurements
({\it e.g.}, in $J/\psi\,\pi^0$ and $Dh^0$)
that help to ensure that the theoretical uncertainty is under control.
LHCb plans to measure the angle $\gamma$ with a precision
of $2$--$3^\circ$.
A SFF is likely to be able to improve this precision to about $1^\circ$. 
LHCb can make a precise measurement of the zero of the
forward-backward asymmetry in $B^0 \to K^{*0}\mu^+\mu^-$,
but a SFF  can also measure the inclusive channel $b \to s \ell^+\ell^-$,
which is theoretically a significantly cleaner observable~\cite{Ghinculov:2003qd}. 

The broad program of a SFF thus provides a very comprehensive set
of measurements, extending what will already have been achieved
by LHCb at that time.  This will be of great importance for the study of
flavour physics in the LHC era and beyond.

\newpage

\section{Phenomenological Impact}
\begin{figure}[tb!]
  \begin{center}
    \includegraphics[width=0.45\textwidth]{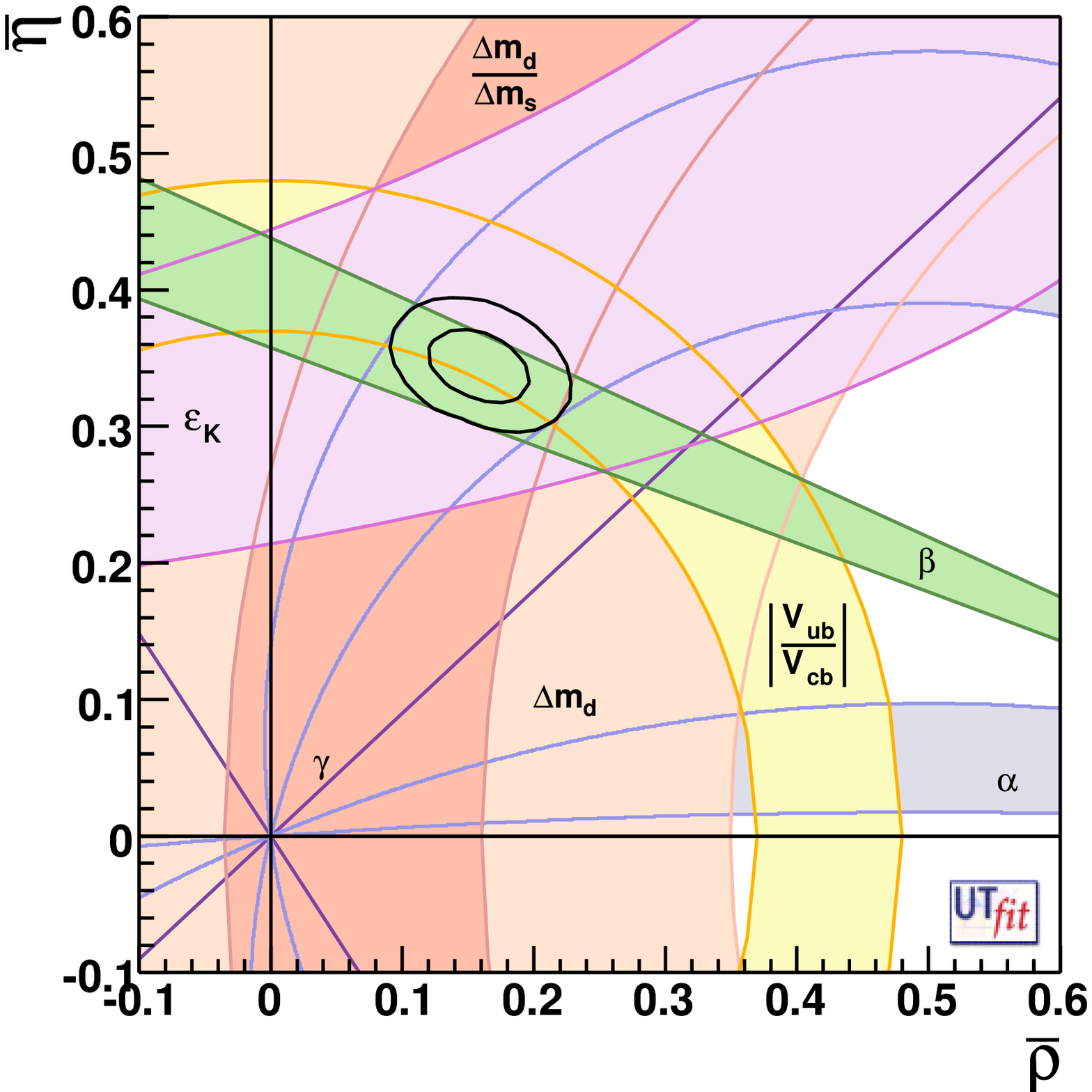}
    \includegraphics[width=0.45\textwidth]{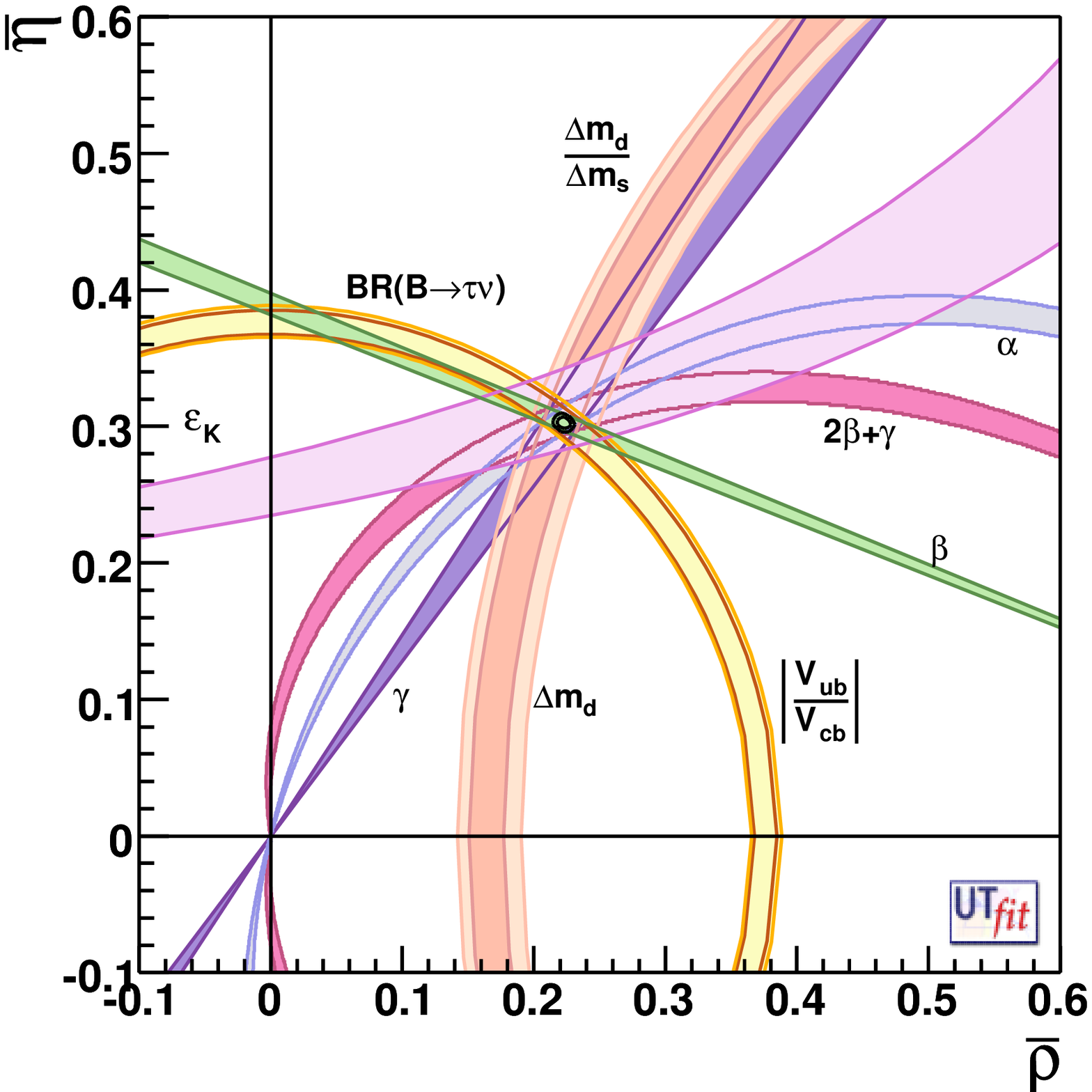}
    \caption{
      Regions corresponding to $95\%$ probability for the CKM
      parameters $\bar\rho$
      and $\bar\eta$ selected by different constraints, assuming
      present central values with present errors (left) or with
      errors expected at a  SFF tuning central values to have
      compatible constraints (right).
    }
    \label{fig:ckm}
  \end{center}
\end{figure}
The power of a SFF to observe NP and to determine the 
CKM parameters precisely is manifold.
In the following, we present a few highlights  of  the phenomenological
impact (for more detailed analyses see~\cite{Hashimoto:2004sm,Bona:2007qt,Akeroyd:2004mj,Hewett:2004tv,superKEKB}).

\begin{figure}[htb!]
  \begin{center}
    \includegraphics[width=0.45\textwidth]{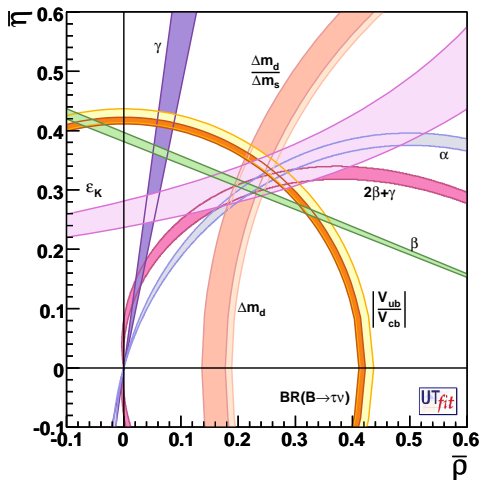}
    \caption{Region corresponding to 95\% probability for the CKM parameter
   $\rhobar$ and $\etabar$ selected by the different constraints, assuming
   todays central values with the precision of a SFF. Note for example
that the band corresponding to the $\gamma$  measurement does not pass through
the intersection of other constraints.}\label{fig:smfit}
  \end{center}
\end{figure}

{\it Precise Determination of CKM Parameters in the SM:} Most of 
the measurements described in the previous section can be used to select a 
region in the $\rhobar$--$\etabar$ plane as shown in Figure~\ref{fig:ckm}.
The corresponding numerical results are given in Table~\ref{tab:smfit}.
The results indicate that a precision of a fraction of a percent can be 
reached, significantly improving the current situation, and providing 
a generic test of the presence of NP at that level of precision. 
Note that in the right plot of  Figure~\ref{fig:ckm} -  
where the  expected precision  offered by a SFF is used -  
the validity of the SM is assumed, so the compatibility of all constraints 
is put in by hand.
In contrast, in Figure~\ref{fig:smfit} we assume that all results take
the central values of their current world averages
with the expected precision of a SFF.
In this case, the hints of discrepancies present in today's data
 have evolved into fully fledged NP discoveries.

\begin{table}[ht]
  \caption{
    Uncertainties of the CKM parameters obtained from the Standard Model fit
    using the experimental and theoretical information available today (left)
    and at the time of a SFF (right). The precision corresponds to the plots 
    in  Figures~\ref{fig:ckm} and  \ref{fig:smfit}.}
  \begin{center}
    \begin{tabular}{lll}
      \hline\hline
      Parameter               &   SM Fit today        &  SM Fit at a SFF \\
      \hline
      $\overline {\rho}$      &   $0.163\pm 0.028$    &  $\pm 0.0028$      \\
      $\overline {\eta}$      &   $0.344\pm 0.016$    &  $\pm 0.0024$      \\
      $\alpha$ ($^{\circ}$)   &   $92.7\pm4.2$        &  $\pm 0.45$        \\
      $\beta$ ($^{\circ}$)    &   $22.2\pm0.9$        &  $\pm 0.17$        \\
      $\gamma$ ($^{\circ}$)   &   $64.6\pm4.2$        &  $\pm 0.38$        \\
      \hline
    \end{tabular}
  \end{center}
  \label{tab:smfit}
\end{table}
Of course,  many of the measurements used for the SM  determination of
$\rhobar$--$\etabar$ can be affected by the presence of NP.
Thus, unambiguous NP searches require a determination of $\rhobar$  and $\etabar$
in the presence of arbitrary NP contributions, which can be done using $\Delta F=2$ processes.

{\it New Physics in Models with Minimal Flavour Violation:}
The basic assumption of Minimal Flavour Violation (MFV)~\cite{Gabrielli:1994ff,Buras:2000dm,D'Ambrosio:2002ex} is that 
NP does not introduce new sources of flavour and $\CP$ violation.
Hence the only flavour-violating couplings are the SM Yukawa couplings.
One can assume that the top Yukawa coupling is dominant
in the simplest case with one Higgs doublet and - with some exceptions - also
in the case with two Higgs doublets with small $\tan \beta$; this means
that all NP effects amount to a real contribution added to the SM
loop function generated by virtual top exchange. In particular, in the $\Delta B = 2$ amplitude,
MFV NP may be parameterized as $$S_0(x_t) \to S_0(x_t) + \delta S_0$$
where the function $S_0(x_t)$ represents the top contribution
in the box diagrams and $\delta S_0$ is the NP contribution.
Therefore, in this class of MFV models, the NP contribution to 
all $\Delta F = 2$ processes is universal,
and the effective Hamiltonian retains the SM  structure.

\begin{figure}[t]
  \begin{center}
    \includegraphics[width=0.45\textwidth]{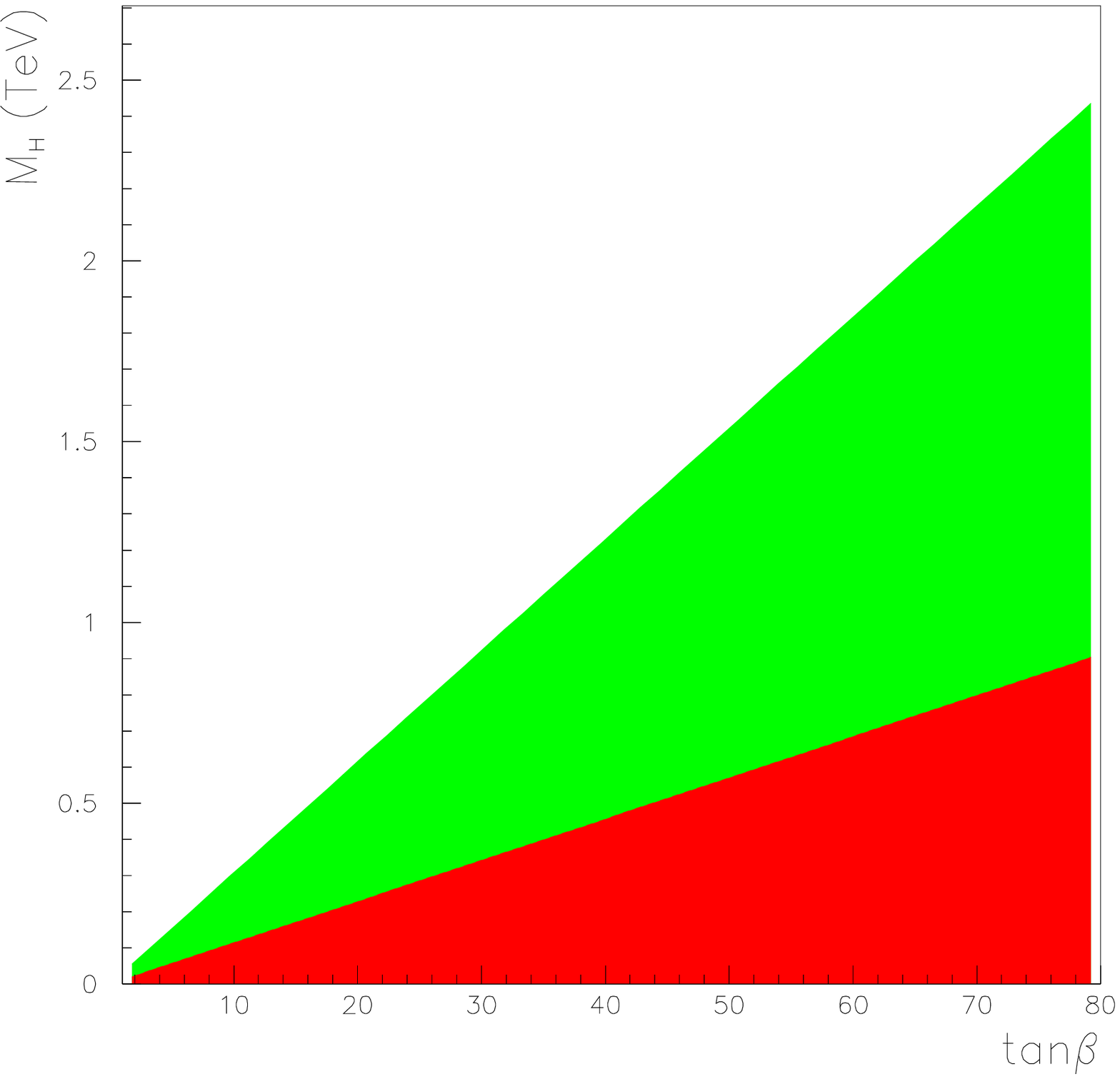} 
    \includegraphics[width=0.45\textwidth]{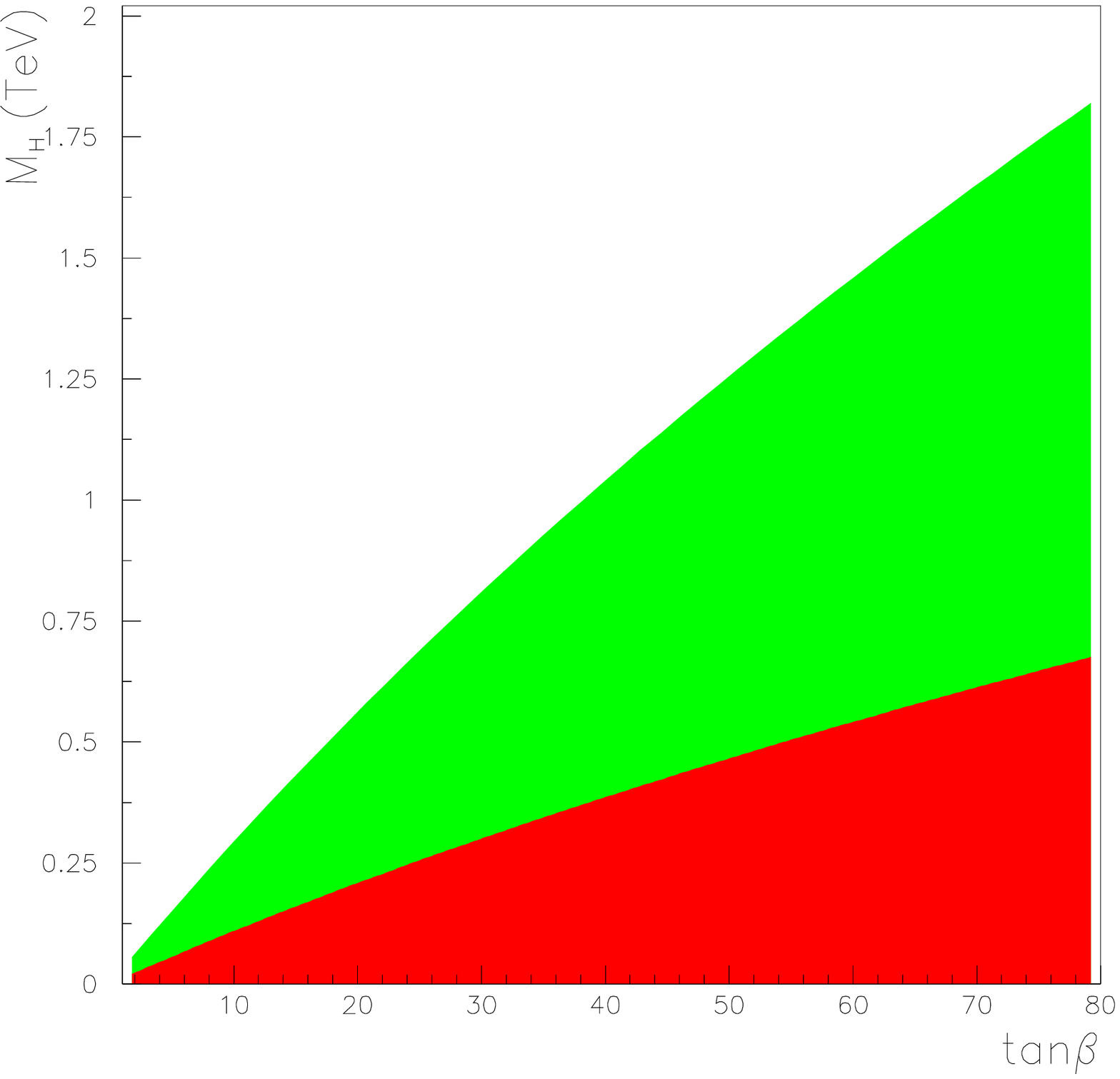} 
    \caption{
      Exclusion regions at 95\% probability 
      in the $M_{H^\pm}$--$\tan \beta$ plane for the 2HDM-II (left) 
      and the MSSM (right) obtained
      assuming the Standard Model value of ${\cal B}(B\to\ell\nu)$ measured 
      with $2 \ {\rm ab}^{-1}$ (dark (red) area) and
      $75 \ {\rm ab}^{-1}$ (dark (red) + light (green) area). 
      In the MSSM case, we have used
      $\epsilon_0 \sim 10^{-2}$~\cite{Isidori:2007zm}.
    }
    \label{fig:btaunu}
  \end{center}
\end{figure}

Following Ref.~\cite{D'Ambrosio:2002ex}, this value can be converted into a NP scale using
\begin{equation}
  \delta S_0 =  4 a \left( \frac{\Lambda_0}{\Lambda}\right)^2\,,
\end{equation}
where $\Lambda_0=Y_t \sin^2 \theta_W M_W/\alpha \approx 2.4 \ {\rm TeV}$
is the SM  scale, $Y_t$ is the top Yukawa coupling,
$\Lambda$ is the NP  scale and $a$ is an unknown (but real)
Wilson coefficient of ${\cal O}(1)$.

The UT analysis can constrain the value of the NP  parameter $\delta S_0$ together
with $\rhobar$ and $\etabar$.
In the absence of a NP  signal, $\delta S_0$ is distributed around zero.
From this distribution, we can obtain a lower bound on the NP scale $\Lambda$.

For a one-Higgs-doublet model (1HDM) or a two-Higgs-doublet model (2HDM) in the low 
$\tan \beta$ regime, the combination of measurements at a SFF 
and the improved lattice 
results give
\begin{equation}
  \Lambda > 14 \ {\rm TeV}~@~95\% \ {\rm CL}
\end{equation}

\begin{figure}[t]
  \begin{center}
    \includegraphics[width=0.45\textwidth]{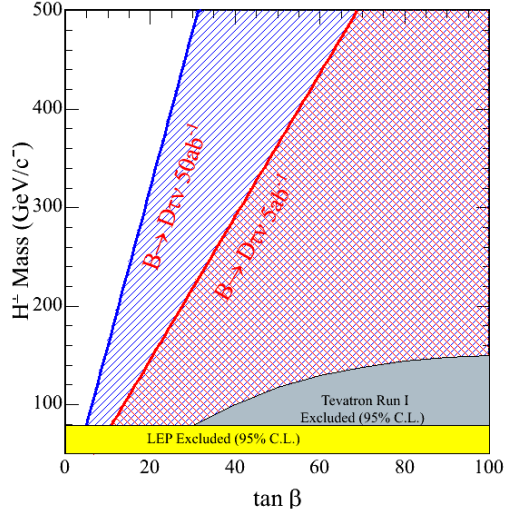}
    \caption{
      Exclusion region 
      in the $M_{H^\pm}$--$\tan \beta$ plane 
      assuming the SM value of ${\cal B}(B\to D\ell\nu)$ measured with
      $5 \ {\rm ab}^{-1}$ and with $50 \ {\rm ab}^{-1}$.}
    \label{fig:btaunu2}
  \end{center}
\end{figure}

These bounds are a factor of three larger than those available 
today~\cite{Bona:2005eu}.
This means that even in the ``worst case'' scenario,
{\it i.e.},~in models with MFV at small $\tan \beta$,
the sensitivity of flavour-violating processes to NP 
is strong enough to allow for the study of the flavour-violating
couplings of new particles with masses up to $600$ GeV.
This conversion to a NP scale in the MFV case deserves 
further explanation.
Consider that the SM reference scale
corresponds to virtual $W$-exchange in the loops. 
As MFV has the same flavour violating couplings as the SM, the
MFV-NP scale is simply translated to a new virtual particle
mass as $\Lambda/\Lambda_0\times  M_W$.
It must be noted, however, that as soon as one considers large $\tan \beta$,
or relaxes the MFV assumption in this kind of analysis,
the NP scale is raised by at least a factor of three,
covering the whole range of masses accessible at the LHC.
In fact the RGE-enhanced contribution of the scalar operators
(absent or subleading in the small $\tan \beta$ MFV case)
typically sets bounds an order of magnitude stronger than those
on the SM current-current operator,
correspondingly increasing the lower bound on the NP scale.
This is the case, for instance, in the Next-to-Minimal Flavour Models (NMFV)  
discussed in Ref.~\cite{Agashe:2005hk} as described in 
the analysis of Ref.~\cite{Bona:2007vi}.

The large $\tan \beta$ scenario offers additional opportunities
to reveal NP by enhancing flavour-violating couplings in
$\Delta B = 1$ processes with virtual Higgs exchange.
This can be the case in decays such as $B\to\ell\nu$ or $B\to D\tau\nu$
whose branching ratios are strongly affected by a charged Higgs
for large values of $\tan\beta$.
In Figure~\ref{fig:btaunu} we show the region excluded in the
$M_{H^\pm}$--$\tan \beta$ plane by the measurement of ${\cal B}(B\to\ell\nu)$
with the precision expected at the end of the current $B$~Factories
and at a SFF, assuming the central value given by the SM.
It is apparent that a SFF  pushes the lower bound on $M_{H^\pm}$,
corresponding, for example, to $\tan \beta \sim 50$ from the hundreds 
of GeV region up to about 2 TeV, both in the 2HDM-II and in the MSSM.
Another interesting possibility is 
to test lepton flavour universality by measuring the ratio
$R_B^{\mu/\tau} = {\cal B}(B\to\mu\nu)/{\cal B}(B\to\tau\nu)$,
which could have a ${\cal O}(10\%)$ deviation from its SM value
at large $\tan \beta$~\cite{Isidori:2006pk,Masiero:2005wr},
whereas the relative error on the individual branching fraction measurements
at a SFF  is expected to be $5\%$ or less. 
In Figure \ref{fig:btaunu2} we show the region excluded in the
$M_{H^\pm}$--$\tan \beta$ plane by the measurement of ${\cal B}(B\to D\ell\nu)$
at a SFF, assuming the central value given by the SM.

{\it MSSM with Generic Squark Mass Matrices:}
\begin{figure}[t]
  \begin{center}
    \begin{tabular}{cc}
      \includegraphics[width=0.46\textwidth]{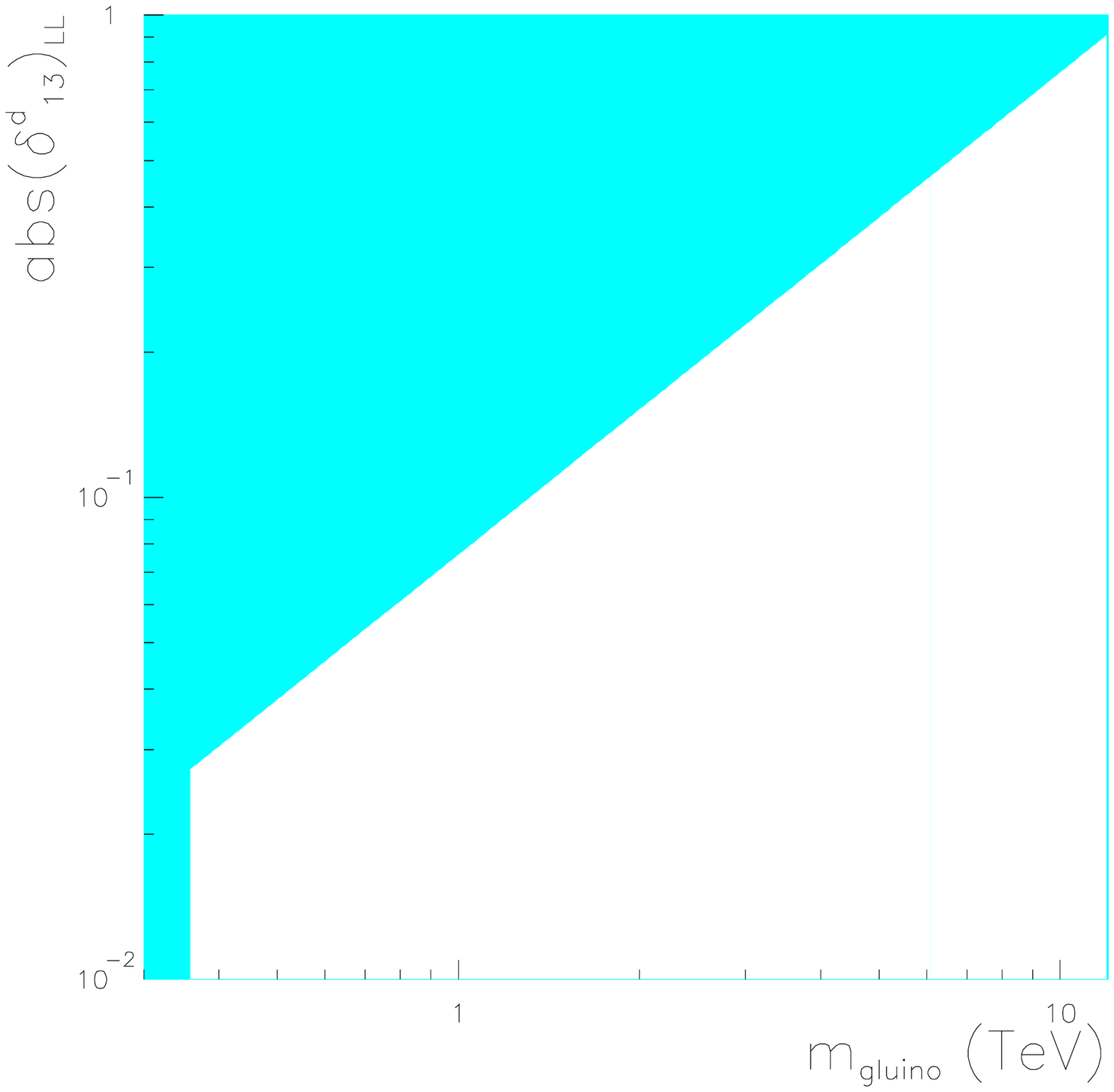} &
      \includegraphics[width=0.46\textwidth]{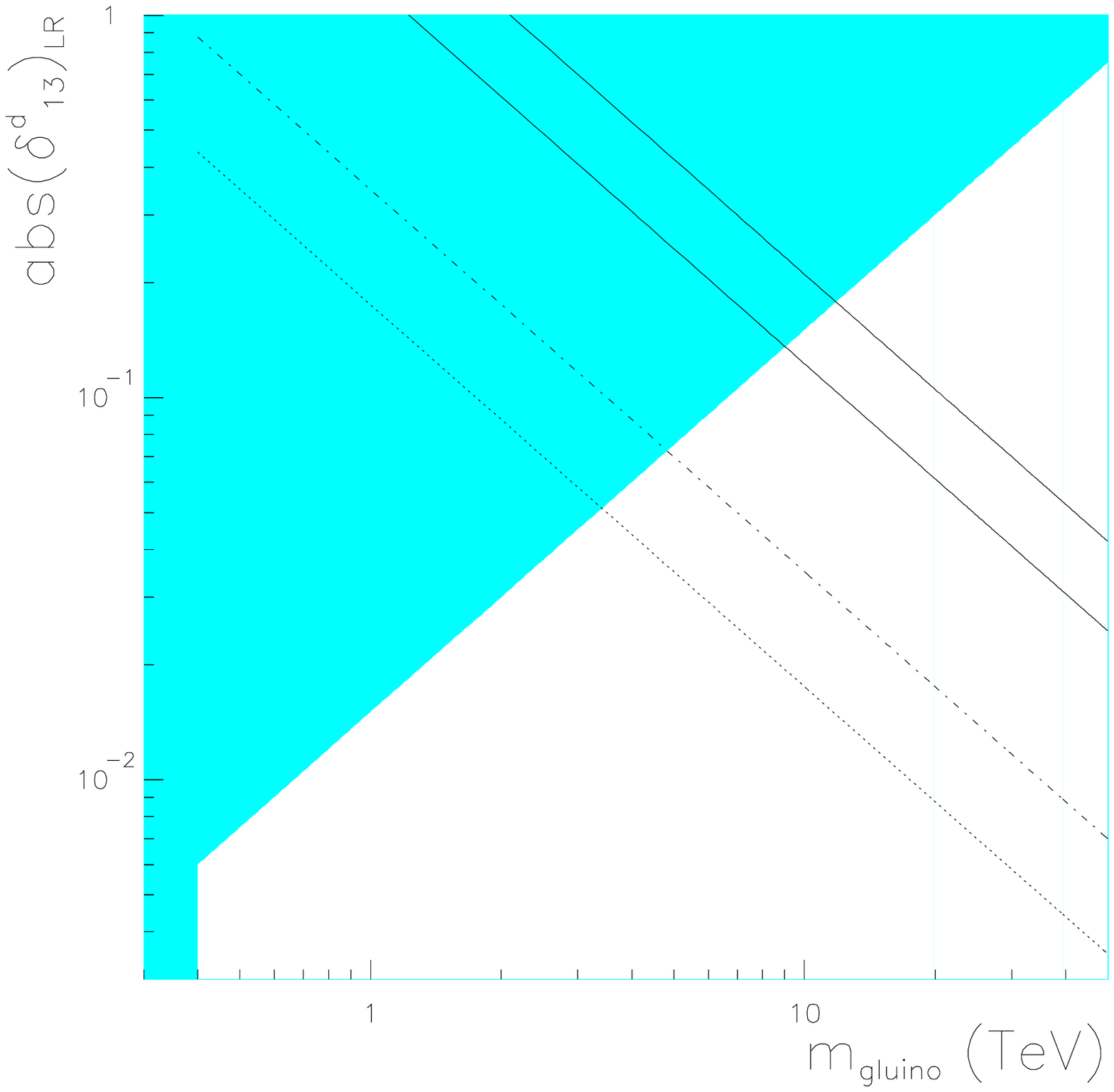} \\
      \includegraphics[width=0.46\textwidth]{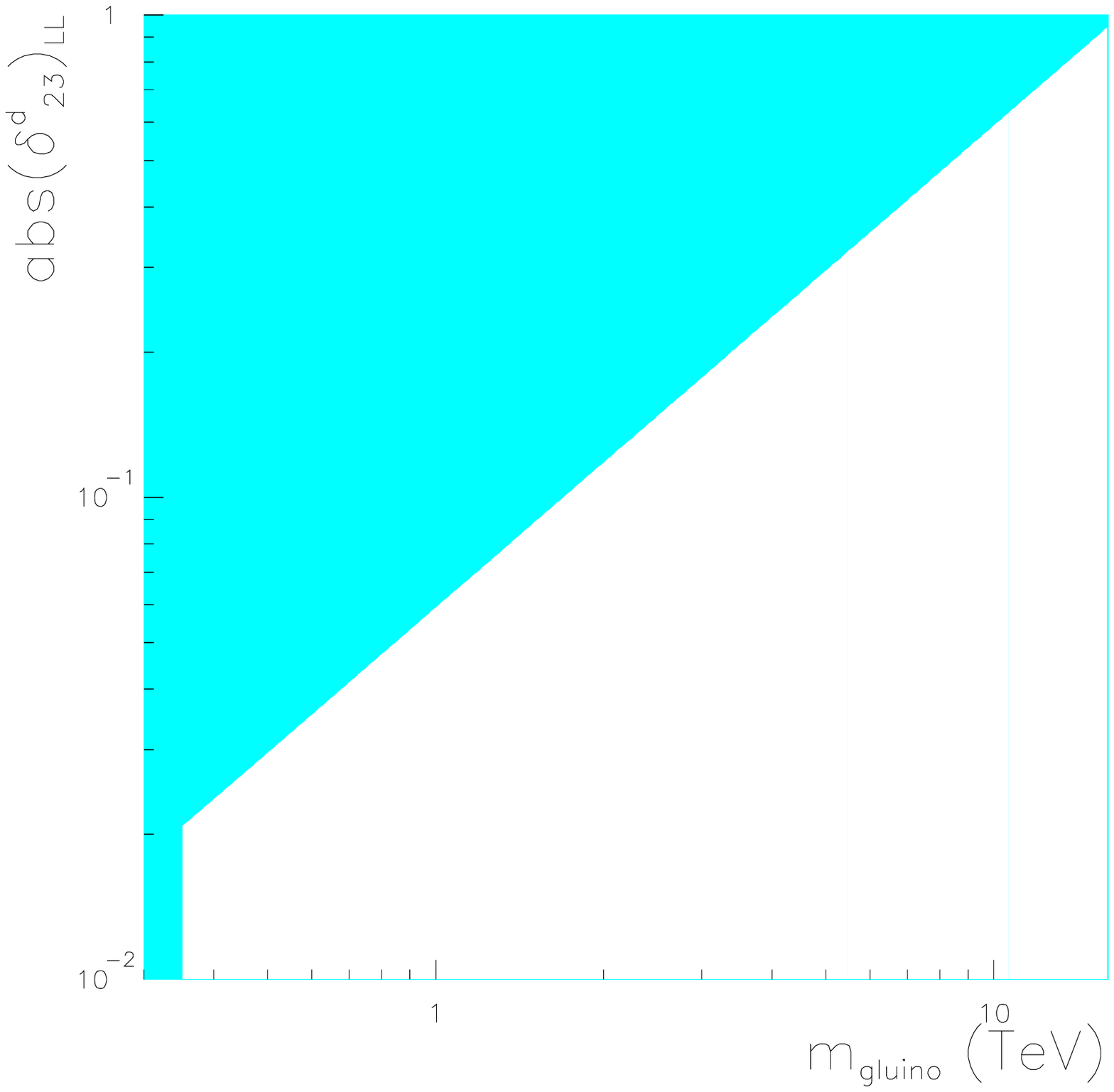} &
      \includegraphics[width=0.46\textwidth]{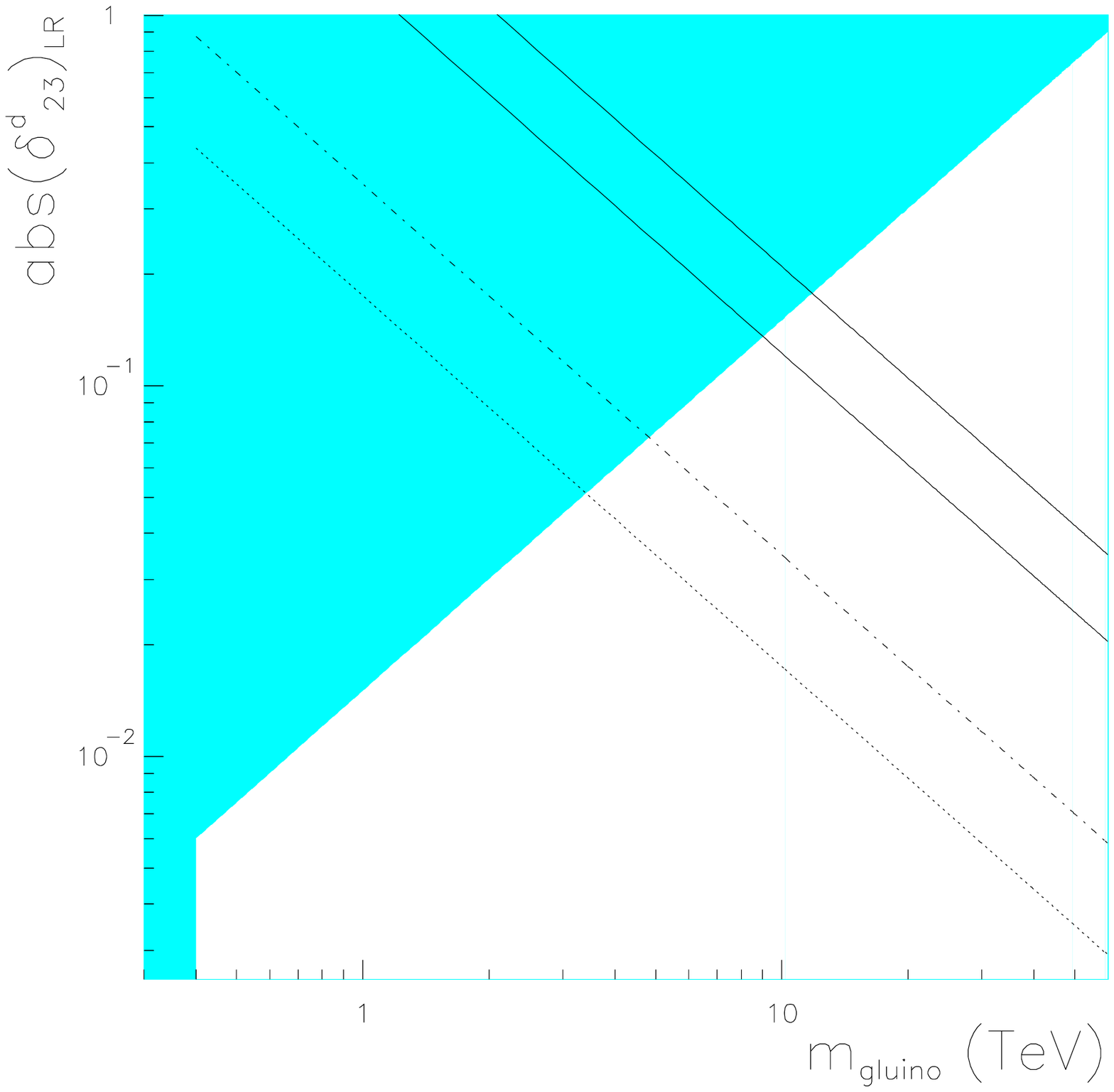}
    \end{tabular}
    \caption{
      Sensitivity region of SFF in the
      $m_{\tilde g}$--$\vert(\delta^d_{ij})_{AB}\vert$ plane.
      The region is obtained by requiring that the reconstructed
      MI is $3\sigma$ away from zero.
      The cases of $(\delta^d_{13})_{LL}$ (upper left),
      $(\delta^d_{13})_{LR}$ (upper right),
      $(\delta^d_{23})_{LL}$ (lower left) and
      $(\delta^d_{23})_{LR}$ (lower right) are shown.
      For LR MIs the theoretical upper bound (allowed parameter region is below these lines)
      discussed in the text is also shown for $\tan \beta= 5, 10, 35, 60$
      (dashed, dotted, dot-dashed, solid line respectively).
    }
    \label{fig:MIvsMg}
  \end{center}
\end{figure}
There is also an impressive  impact of a SFF  on
the parameters of the MSSM with generic squark mass matrices
parameterized using the mass insertion (MI) approximation~\cite{Hall:1985dx}.
In this framework, the NP  flavour-violating couplings are the complex MIs.
For simplicity, we consider only the dominant gluino contribution.
The relevant parameters are therefore the gluino mass $m_{\tilde g}$,
the average squark mass $m_{\tilde q}$ and the MIs $(\delta^{d}_{ij})_{AB}$,
where $i,j=1,2,3$ are the generation indices and
$A,B=L,R$ are the labels referring to the helicity of the SUSY partner quarks.
For example, the parameters relevant to $b \to s$ transitions are
the two SUSY masses and the four MIs $(\delta^{d}_{23})_{LL,LR,RL,RR}$.
In order to simplify the analysis, we consider the
contribution of one MI at a time.
This is justified to some extent by the hierarchy of
the present bounds on the MIs.
In addition, barring accidental cancellations,
the contributions from two or more MIs would produce larger NP effects
and therefore make the detection of NP  easier,
while simultaneously making the phenomenological analysis more 
involved~\cite{Borzumati:1999qt,Besmer:2001cj}.
\begin{figure}[t]
  \begin{center}
    \begin{tabular}{cc}
     \multicolumn{2}{c}{
      \includegraphics[width=0.65\textwidth]{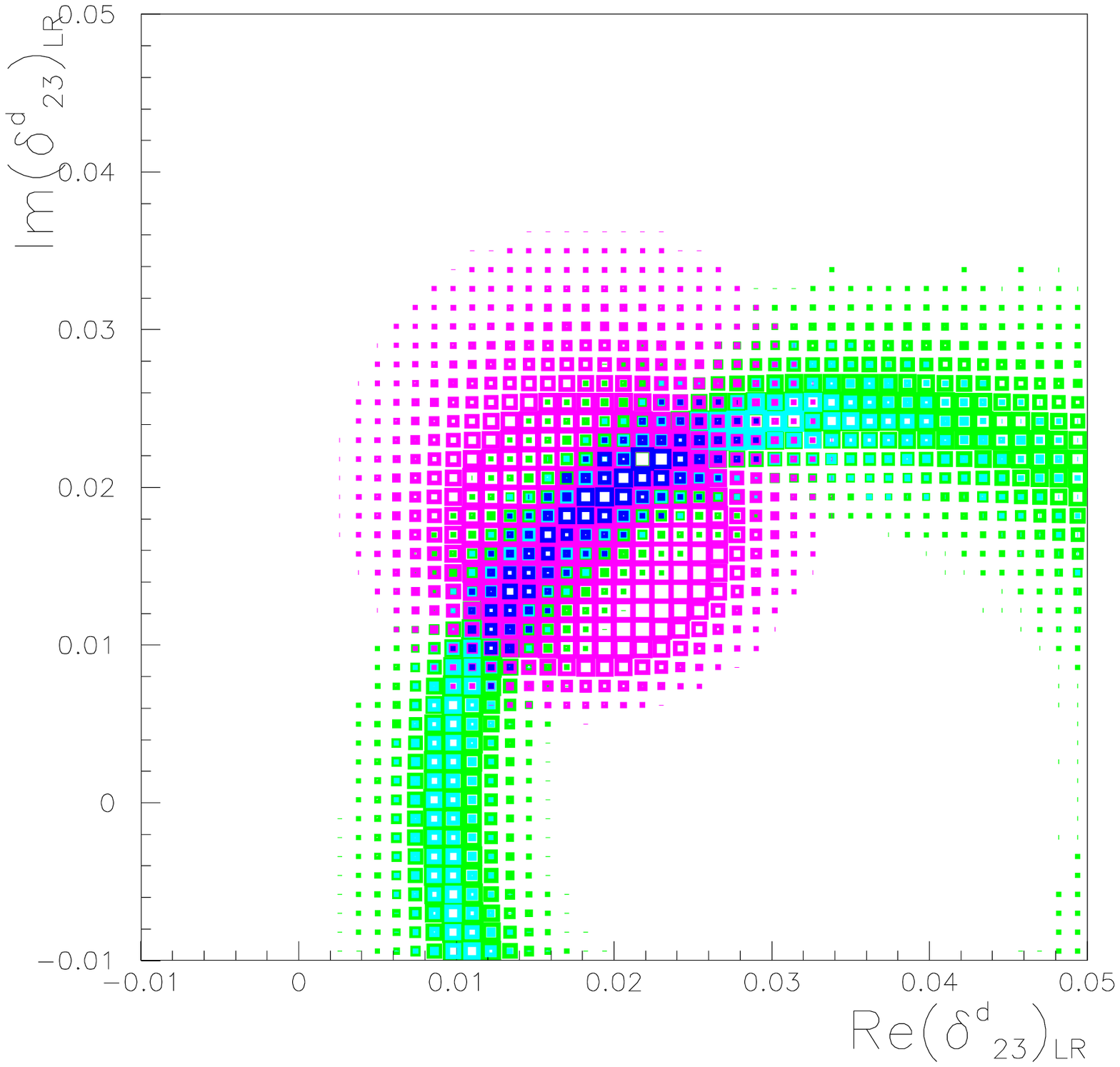}} \\
    \end{tabular}
    \caption{
      Density plot of the region in the
      $\Re(\delta^d_{23})_{LR}$--$\Im(\delta^d_{23})_{LR}$ for
      $m_{\tilde q}=m_{\tilde g}=1 \ \mathrm{TeV}$ generated
      using SFF measurements. Different colours correspond to different
      constraints: $\BR(B\to X_s\gamma)$ (green),
      $\BR(B\to X_s\ell^+\ell^-)$ (cyan), $A_{CP}(B\to X_s\gamma)$ (magenta),
      all together (blue). Central values of constraints corresponds to
      assuming $(\delta^d_{13})_{LL}=0.028 e^{i\pi/4}$.
    }
    \label{fig:MI23LR}
  \end{center}
\end{figure}
The analysis presented here  is based on results and techniques
developed in Refs.~\cite{Gabbiani:1996hi,Becirevic:2001jj,Ciuchini:2002uv}.
The aim of this analysis is twofold.
On the one hand, we want to show the bounds on the MSSM parameter space
as they would appear at a SFF.
For this purpose, we first simulate the signals produced by the MSSM
for a given value of one MI.
We then check how well we are able to determine this value
using the constraints coming from a SFF.
In particular, we examine  the ranges of masses
and MIs for which clear NP  evidence,
given by a non-vanishing value of the extracted MI, can be obtained.
In Figure~\ref{fig:MIvsMg} we show for some of the different MIs,
the observation region in the plane $m_{\tilde g}$--$\vert\delta^d\vert$
obtained by requiring that the absolute value of the reconstructed MI
is more than $3\sigma$ away from zero.
For simplicity we have taken $m_{\tilde q}\sim m_{\tilde g}$.
From these plots, one can see that  a SFF  could detect NP  effects
caused by SUSY masses up to $10$--$15$ TeV
corresponding to $(\delta^d_{13,23})_{LL}\sim 1$.
Even larger scales could be reached by $LR$ MIs.
However overly large LR MIs are known to produce
charge- and colour-breaking minima in the MSSM potential~\cite{Casas:1996de},
which can be avoided by imposing the bounds
shown in the LR plots of Figure~\ref{fig:MIvsMg}.
These bounds decrease as $1/m_{\tilde q}$ and
increase linearly with $\tan \beta$.
Taking them into account,
we can see that still LR MIs are sensitive to gluino masses
up to $5$--$10$ TeV for $\tan \beta$ between 5 and 60.
The plots of Figure~\ref{fig:MIvsMg} show the values of the MI
that can be reconstructed if SUSY masses are below $1$ TeV.
In the cases considered we find
$(\delta^{d}_{13})_{LL}=2$--$ 5\times 10^{-2}$,
$(\delta^{d}_{13})_{LR}=2$--$15\times 10^{-3}$,
$(\delta^{d}_{23})_{LL}=2$--$ 5\times 10^{-1}$ and
$(\delta^{d}_{23})_{LR}=5$--$10\times 10^{-3}$.
These value are typically one order of magnitude smaller than
the present upper bounds on the MIs~\cite{silvckm}.

Figure~\ref{fig:MI23LR} shows a
simulation of how well the the mass insertions (MIs), related to the
off-diagonal entries of the squark mass matrices, could be reconstructed 
at a  SFF. Figure \ref{fig:MI23LR} displays 
the allowed region in the plane
$\Re(\delta^d_{ij})_{AB}$--$\Im(\delta^d_{ij})_{AB}$
with a value of $(\delta^d_{ij})_{AB}$
allowed from the present upper bound,
$m_{\tilde g}=1$ TeV and using the SFF measurements as constraints.
The relevant constraints come from
$\BR(b\to s\gamma)$, $A_{\CP}(b\to s\gamma)$, $\BR(b\to s\ell^+\ell^-)$,
$A_{\CP}(b\to s\ell^+\ell^-)$, $\Delta m_{B_s}$ and $A^s_{\rm SL}$.
It is apparent the key role of $A_{\CP}(b\to s\gamma)$
together with the branching ratios of $b\to s\gamma$ and $b\to s\ell^+\ell^-$.
The zero of the forward-backward asymmetry
in $b\to s\ell^+\ell^-$, missing in the present analysis,
is expected to give an additional strong constraint,
further improving the already excellent extraction of
$(\delta^d_{23})_{LR}$ shown in Figure~\ref{fig:MI23LR}.

{\it Lepton Flavour Violation in $\tau$ Decays:}   
The search for Flavour Changing Neutral Current (FCNC) transitions
of charged leptons is one of the most promising directions
to search for physics beyond the SM. In the last 
few years neutrino physics has provided unambiguous indications 
about the non-conservation of lepton flavour, 
we therefore expect this phenomenon to occur also in the charged lepton sector.
FCNC transitions of charged leptons could occur well beyond any 
realistic experimental resolution if the light neutrino mass matrix ($m_{\nu}$)
were the only source of Lepton Flavour Violation (LFV). 
However, in many realistic extensions of the SM this is not the case. 
In particular, the overall size of $m_{\nu}$
is naturally explained by a strong suppression associated to the
breaking of the total Lepton Number (LN), 
which is not directly related to the size of  LFV interactions. 

Rare FCNC decays of the $\tau$ lepton are particularly interesting
since the LFV sources involving the third generation are naturally
the largest. In particular, searches of  $\tau \to \mu \gamma$
at the $10^{-8}$ level or below are extremely interesting even
taking into account the present stringent bounds on
$\mu \to e \gamma$.  We illustrate this with one example where the
comparison of possible bounds on (or evidences for)
$\tau \to \mu \gamma$, $\mu \to e \gamma$ and other LFV rare decays 
provides a unique tool to identify the nature of the NP  model.

\begin{figure}
  \centering
  \includegraphics[angle=-90, width=0.70\textwidth]{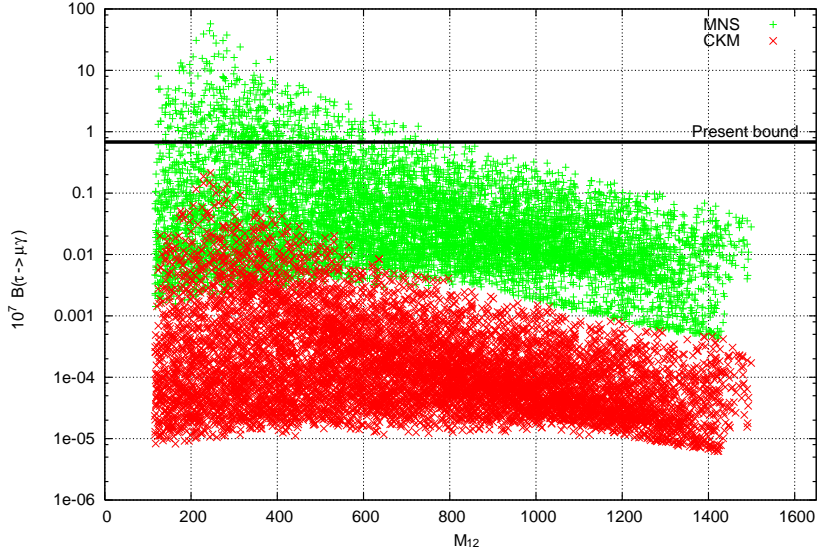}
  \caption{
    $\BR(\tau\to \mu\gamma)$ in units of $10^{-7}$ {\it vs.}
    the high energy universal gaugino mass ($M_{1/2}$)
    within a $SO(10)$ framework~\cite{Calibbi:2006nq}.
    The plot is obtained by scanning the LHC accessible parameter space
    $m_0\leq 5 \ {\rm TeV}$ for $\tan \beta = 40$.
    Green or light (red or dark) points correspond to the PMNS (CKM) case,
    namely the scenario where $Y_{\nu} = U_{\rm PMNS}$
    ($Y_{\nu} = V_{\rm CKM}$).
    The thick horizontal line denotes the present experimental sensitivity.
The expected  SFF sensitivity is $2\times 10^{-9}$.
  }
  \label{masieroLFV06}
\end{figure}

In Figure~\ref{masieroLFV06}, we show the prediction
for $\BR(\tau\to\mu\gamma)$ within a SUSY SO(10) framework
for the accessible LHC SUSY parameter space $M_{1/2}\leq 1.5 \ {\rm TeV}$,
$m_0 \leq 5 \ {\rm TeV}$ and $\tan \beta = 40$~\cite{Calibbi:2006nq}.
Note that the scenarios where the source of LFV violation is governed by 
neutrino mass matrix $Y_{\nu}=U_{\rm PMNS}$ and where $Y_{\nu} = V_{\rm CKM}$ 
can be distinguished by the measurement of $\BR(\tau \to \mu\gamma)$ at a SFF.

{\it Little Higgs Models:} 
These models address the tension 
between the naturalness of the electroweak scale and the precision electroweak 
measurements showing no evidence for new physics up to $5-10$ TeV.
The Littlest Higgs model~\cite{Arkani-Hamed:2002qy}
is based on a $SU(5)/SO(5)$ non-linear sigma model. 
It is strongly 
constrained by the electroweak precision data due to tree-level 
contributions of the new particles.

\begin{table}
  \caption{Upper bounds on some LFV decay branching ratios in the LHT model 
    with a new physics scale $f = 500 \ {\rm GeV}$, 
    after imposing constraints on 
    $\mu^-\to e^-\gamma$, $\mu^-\to e^-e^+e^-$, $\tau^- \to \mu^- \pi^0$ 
    and $\tau^- \to e^- \pi^0$. 
    \label{tab:bounds}
  }
  \begin{center}
    \begin{tabular}{|c|c|c|c|}
      \hline
      Decay & Upper bound \\
      \hline\hline
      $\tau^- \to e^-\gamma$       & $1   \cdot 10^{- 8}$ \\
      $\tau^- \to \mu^-\gamma$     & $2   \cdot 10^{- 8}$ \\
      $\tau^- \to e^-e^+e^-$       & $2   \cdot 10^{- 8}$ \\
      $\tau^- \to \mu^-\mu^+\mu^-$ & $3   \cdot 10^{- 8}$ \\
      \hline
    \end{tabular}
  \end{center}
\end{table}

Implementing an additional discrete symmetry, 
so-called T-parity~\cite{Cheng:2003ju}, 
constrains the new particles to contribute at the loop-level only 
and allows for a NP scale around $500 \ {\rm GeV}$. 
It also calls for additional (mirror) 
fermions providing an interesting flavour phenomenology.

The high sensitivity for $\tau$ decays serves as an important tool 
to test the littlest Higgs model with T-parity (LHT), in particular 
to distinguish it from the MSSM~\cite{Blanke:2007db}.
Upper bounds on some lepton flavour violating decay branching ratios
are given in Table~\ref{tab:bounds}.

\begin{table}
  {\renewcommand{\arraystretch}{1.5}
    \caption{
      Comparison of various ratios of branching ratios in the LHT model 
      and in the MSSM without and with significant Higgs contributions.
      \label{tab:ratios}
    }
    \begin{center}
      \begin{tabular}{|c|c|c|c|}
        \hline
        Ratio & \hspace{.8cm} LHT \hspace{.8cm} & MSSM (dipole) & MSSM (Higgs) \\
        \hline\hline
        $\frac{\BR(\mu^-\to e^-e^+e^-)}{\BR(\mu^-\to e^-\gamma)}$  & 
        0.4 -- 2.5 & $\sim6\cdot10^{-3}$ & $\sim6\cdot10^{-3}$ \\
        $\frac{\BR(\tau^-\to e^-e^+e^-)}{\BR(\tau^-\to e^-\gamma)}$ & 
        0.4 -- 2.3 & $\sim1\cdot10^{-2}$ & $\sim1\cdot10^{-2}$ \\
        $\frac{\BR(\tau^-\to \mu^-\mu^+\mu^-)}{\BR(\tau^-\to \mu^-\gamma)}$ &
        0.4 -- 2.3 & $\sim2\cdot10^{-3}$ & $\sim1\cdot10^{-1}$ \\
        \hline 
      \end{tabular}
    \end{center}
    \renewcommand{\arraystretch}{1.0}
  }
\end{table}

By comparison with Table~\ref{tab:superb}, 
these are seen to be well within the reach of a SFF.
However, the large LFV branching ratios are not a specific 
feature of the LHT but a general property  of many new physics models
including the MSSM. 
Nevertheless, as Table~\ref{tab:ratios} clearly shows,  
specific correlations are very suitable to distinguish 
between the LHT and the MSSM. 
The different ratios are a consequence of the fact that in the MSSM 
the dipole operator plays the crucial role in those observables while 
in the LHT the $Z^0$ penguin and the box diagram contributions 
are dominant. 
The pattern is still valid when there is a significant
Higgs contribution in the MSSM, as can be read off 
from Table~\ref{tab:ratios}.

{\it Comparison of different SUSY Breaking Scenarios:} 
In SUSY models the squark and slepton mass matrices
are determined by various SUSY breaking parameters, 
and hence a SFF has the potential to study SUSY breaking scenarios 
through quark and lepton flavour signals. 
This will be particularly important when SUSY particles are found at the LHC,
because flavour off-diagonal terms in these mass matrices 
could carry information on the origin of SUSY breaking and
interactions at high energy scales such as the GUT and the seesaw
neutrino scales. 
Combined with the SUSY mass spectrum obtained at 
energy frontier experiments, it may be possible
to clarify the whole structure of SUSY breaking.
In order to illustrate the potential of a SFF to explore the SUSY
breaking sector, three SUSY models are considered and various 
flavour signals are compared. 
These are $(i)$ the minimal supergravity model (mSUGRA), 
$(ii)$ a SU(5) SUSY GUT model with right-handed neutrinos, 
$(iii)$ the MSSM with U(2) flavour symmetry~\cite{Goto:2003iu}. 
In mSUGRA, the SUSY breaking terms are assumed to be 
flavour-blind at the GUT scale. 
The SU(5) SUSY GUT with right-handed neutrinos is a well-motivated 
SUSY model which can accommodate the gauge coupling unification 
and the seesaw mechanism for neutrino mass generation. There is 
interesting interplay between the quark and lepton sectors in 
this model. Since quarks and leptons are unified in the same GUT
multiplets, quark flavour mixing can be a source of flavour 
mixings in the slepton sector that  induce LFV in the charged 
lepton processes. Furthermore, the neutrino Yukawa coupling 
constants introduce new flavour mixings that are not related to
the CKM matrix. Due to the SU(5) GUT multiplet structure
sizable flavour mixing can occur in the right-handed
sdown sector as well as the left-handed slepton sector, and 
contributions to various LFV and 
quark FCNC processes become large. 
When we require that the neutrino Yukawa coupling 
constants only induce flavour mixing in the 2-3 generation, then the 
constraint from the $\mu \to e \gamma$ process is somewhat relaxed
(so-called non-degenerate case). 
Finally, in the MSSM with U(2) flavour symmetry, 
the first two generations of quarks and squarks are assigned as doublets with
respect to the same U(2) flavour group, whereas those in the third generation 
are singlets. 
Therefore this model explains the suppression  of the FCNC processes 
between the first two generations, but it still provides sizable 
contributions for $b \to s$ transition processes. 

\begin{figure}[t]
  \begin{center}
    \hspace{-0.7cm}\includegraphics[width=0.52\textwidth]{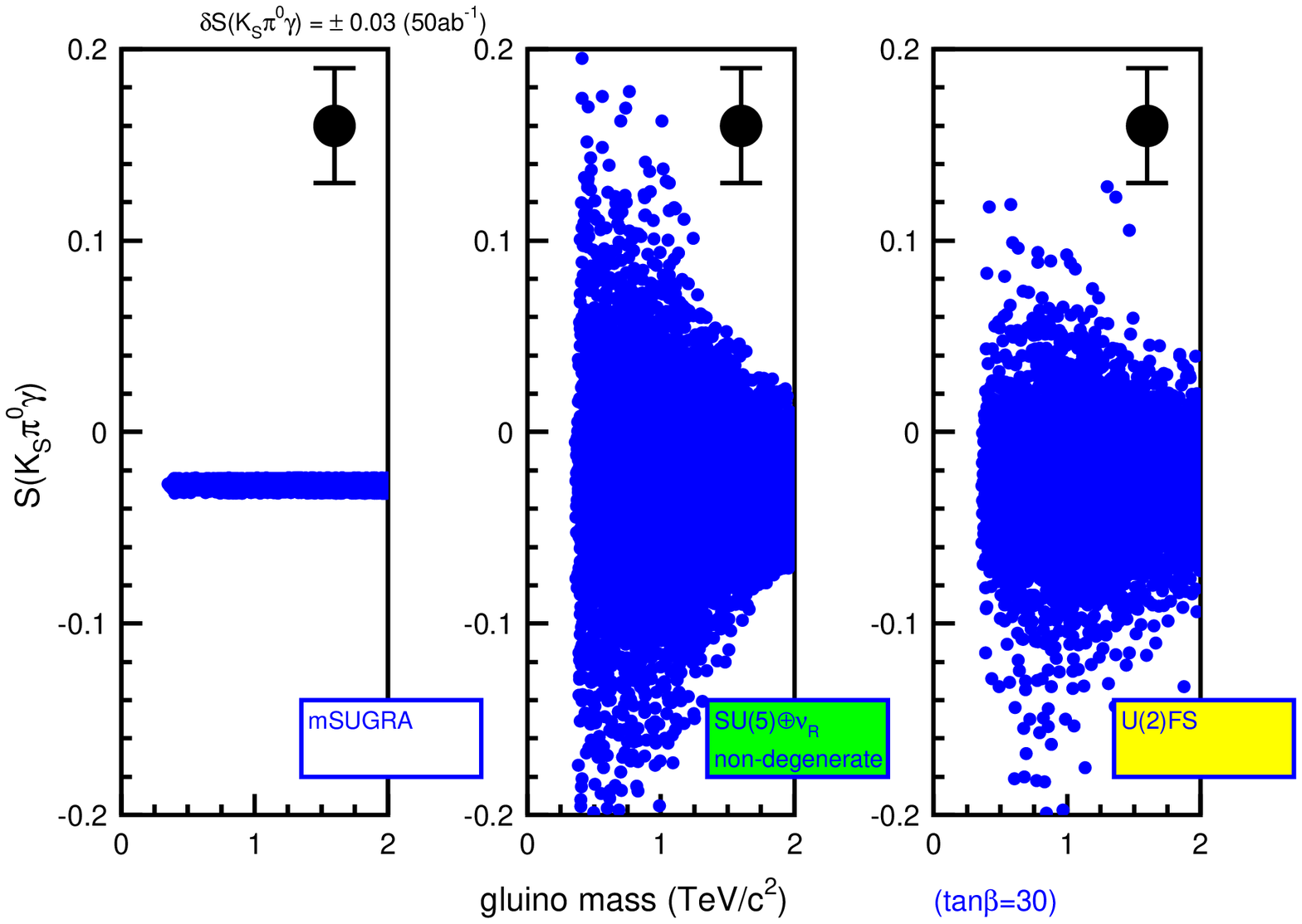}
    \hspace{-0.5cm}\includegraphics[width=0.52\textwidth]{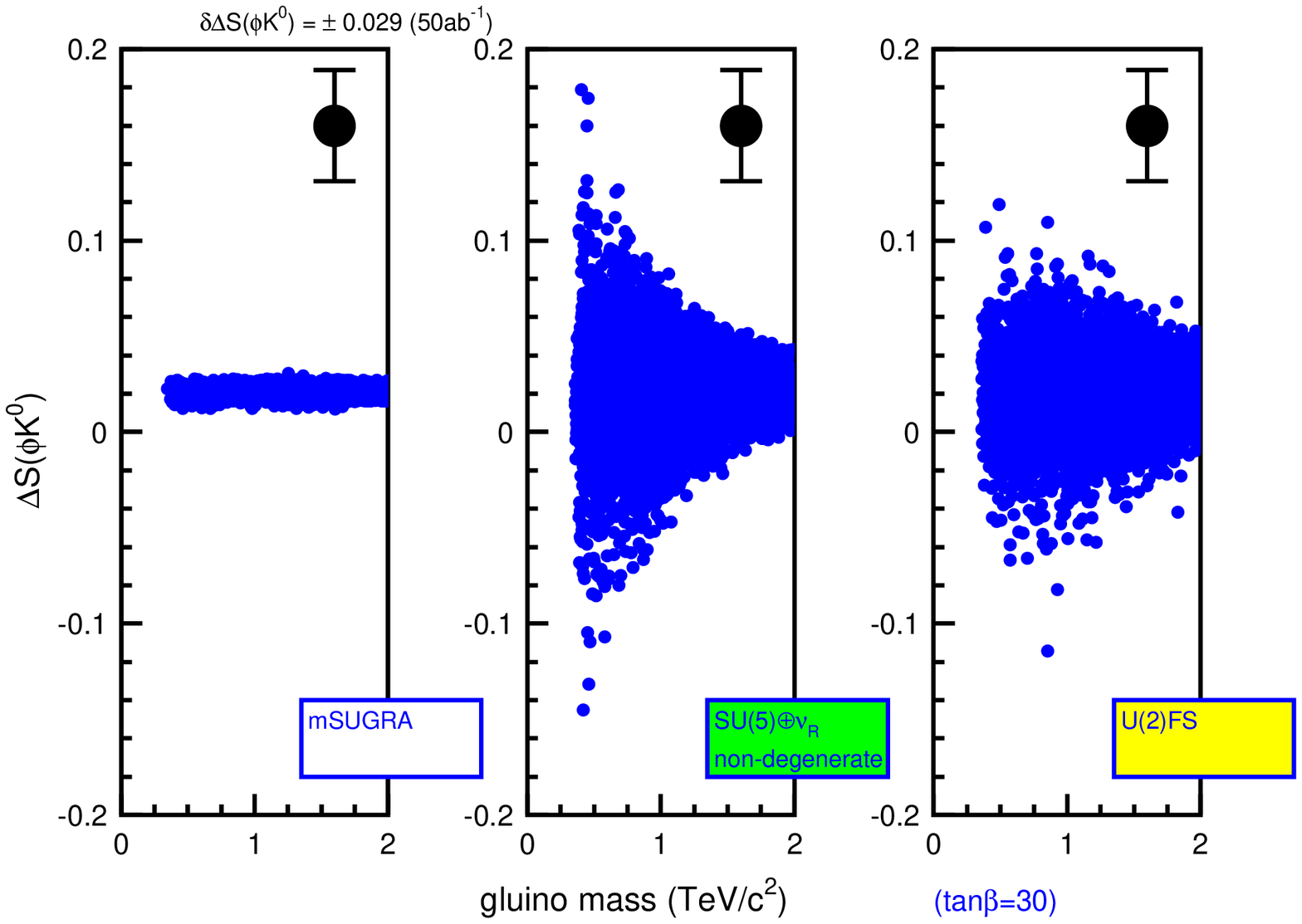}
    \caption{
      Time-dependent asymmetry of $B \to \KS \pi^0 \gamma$
      and the difference between the time-dependent asymmetries of 
      $B \to \phi \KS$ and $B \to J/\psi\,\KS$ modes 
      for three SUSY breaking scenarios:
      mSUGRA(left), SU(5) SUSY GUT with right-handed neutrinos 
      in non-degenerate case (middle), 
      and MSSM with U(2) flavour symmetry (right).  
      The expected SFF sensitivities are also shown.
    }
    \label{Okada1}
  \end{center}
\end{figure}

Flavour signals in the $b \to s$ sector are shown in Figure~\ref{Okada1}
for these three SUSY breaking scenarios.
Scatter plots of
the time-dependent asymmetry of $B \to \KS \pi^0 \gamma$
and the difference between the time-dependent asymmetries of 
$B \to \phi \KS$ and $B \to J/\psi\,\KS$ modes are presented
as a function of the gluino mass. Various phenomenological constraints 
such as ${\cal B}(b \to s \gamma)$, the rate of $B_s$ mixing, 
and neutron and atomic electic dipole moments 
are taken into account as well as SUSY and Higgs particle 
search limits from LEP and TEVATRON experiments. For the SUSY
GUT case, the branching ratios of muon and tau LFV processes 
are also calculated and used to limit the allowed parameter 
space. Sizable deviations can be seen for SU(5) SUSY GUT and
U(2) flavour symmetry cases even if the gluino mass is 1 TeV.
The deviation is large enough to be identified at SFF. On the 
other hand, the deviations are much smaller for the mSUGRA case. 

\begin{figure}[t]
  \begin{center}
    \includegraphics[width=0.55\textwidth]{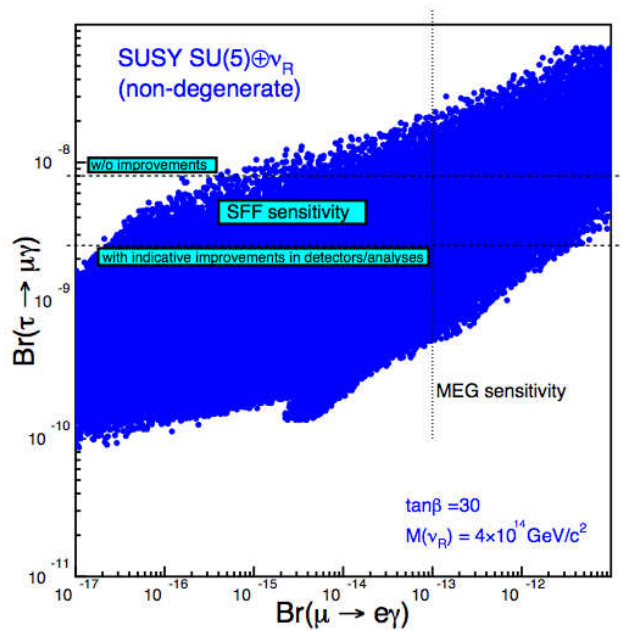}\\
    \caption{
      Correlation between ${\cal B}(\tau \to \mu \gamma)$ and 
      ${\cal B}(\mu \to e \gamma)$  for  SU(5) SUSY GUT 
      with right-handed neutrinos in non-degenerate case. 
      Expected search limits at the SFF for  ${\cal B}(\tau \to \mu \gamma)$
      and for  ${\cal B}(\mu \to e \gamma)$ from MEG are also shown.
    }
    \label{Okada2}
  \end{center}
\end{figure}

The correlation between 
${\cal B}(\tau \to \mu \gamma)$ and ${\cal B}(\mu \to e \gamma)$
is shown in Figure~\ref{Okada2} for the non-degenerate SU(5) SUSY GUT case. 
In this case, both processes can reach current upper bounds. 
It is thus possible that improvements in the $\mu \to e \gamma$ search 
at the MEG experiment and in the $\tau \to \mu \gamma$ search at a SFF 
lead to discoveries of muon and tau LFV processes, respectively.
Notice that the Majorana mass scale that  roughly corresponds to the heaviest 
Majorana neutrino mass is taken to be $M_R = 4 \times 10^{14} \ {\rm GeV}$
in these figures. When the Majorana mass scale is lower, flavour signals
become smaller because the size of the neutrino Yukawa coupling 
constant is proportional to $\sqrt{M_R}$ and LFV branching ratios
scale with $M_R^2$. This means that a SFF can cover some part of the
parameter space from $\tau \to \mu \gamma$ if the Majorana
scale is larger than $10^{13} \ {\rm GeV}$.   
The pattern of LFV signals also depends on the choice of SUSY 
breaking scenarios.
If we take the degenerate case of three heavy Majorana masses in a 
SU(5) SUSY GUT, 
${\cal B}(\mu \to e \gamma)$ can be close to the present experimental
bound while branching ratios of tau LFV processes are generally
less than $10^{-9}$.
The LFV branching ratios for both muon and tau LFV processes are negligible
for the mSUGRA case. In MSSM with U(2) flavour symmetry, LFV signals
depend on how the flavour symmetry is implemented in the lepton sector
so that there is a large model dependence.

\section{Summary}

In conclusion, the physics case of a Super Flavour Factory 
collecting an integrated luminosity of 
$50$--$75$ ab$^{-1}$ is well established.
Many NP sensitive measurements involving $B$ and $D$ mesons and $\tau$ leptons,
unique to a Super Flavour Factory, 
can be performed with excellent sensitivity to new particles with masses up to 
$\sim 100$ (or even $\sim 1000$) TeV. 
The possibility to operate at the $\Upsilon(5{\rm S})$ resonance
makes measurements with $B_s$ mesons also accessible,
and options to run in the tau-charm threshold region and 
possibly with one or two polarized beams further broadens the physics reach.
Flavour- and $\CP$-violating couplings of new particles accessible at the 
LHC can be measured in most scenarios, even in the unfavourable cases
assuming minimal flavour violation.  
Together with the LHC, a Super Flavour Factory could be soon starting the 
project of reconstructing  the NP  Lagrangian. 
Admittedly, this daunting task would be difficult and take many years,
but it provides an exciting objective for accelerator-based 
particle physics in the next decade and beyond.

\section*{Acknowledgements}

We wish to thank Ikaros Bigi and Franz Muheim for useful comments.

\end{document}